\documentclass[journal,inline]{vgtc}  
\usepackage[svgnames]{xcolor}
\definecolor{NavyBlue}{RGB}{0,0,128}

\usepackage{tikz}
\usepackage{xspace}

\usepackage{amsmath}
\usepackage{multirow}
\usepackage{comment}
\usepackage{balance}

\newcommand{\figref}[2]{Fig.~#1 (#2)}
\newcommand{\para}[1]{\vspace{1mm}\noindent\textbf{#1.}}

\newcommand{\participantquote}[1]{\emph{``#1''}}    

\newcommand{\dcdefine}[2]{%
  \expandafter\gdef\csname dctext:#1\endcsname{#2}%
}

\newcommand{\dclabel}[2]{%
  \textbf{{\csname dctext:#1\endcsname}}%
}

\newcommand{\dcref}[1]{%
  \hyperref[fig:desiderata]{\textbf{\csname dctext:#1\endcsname}}%
}

\dcdefine{dc:desiderata}{DC1}
\dcdefine{dc:values}{DC2}
\dcdefine{dc:intuitive}{DC3}
\dcdefine{dc:additional}{DC4}
\dcdefine{dc:llm}{DC5}

\definecolor{HighlightColor}{rgb}{0.85,0.1,0.1}

\newcommand{\revision}[1]{{\color{black}#1}}
\newcommand{\clarification}[1]{{\color{black}#1}}

\onlineid{1204}

\vgtccategory{Research}

\vgtcpapertype{application}

\title{\toolname: An Interactive Visualization System for Biomedical Schema Matching with LLM-Powered Validation}

\author{%
    \authororcid{Eden Wu}{0009-0003-6455-917X}{*}{1}, 
    \authororcid{Dishita G Turakhia}{0000-0002-0200-721X}{*}{1},
    \authororcid{Guande Wu}{0000-0002-9244-173X}{*}{1}, 
    \authororcid{Christos Koutras}{0000-0003-3015-154X}{1},
    \authororcid{Sarah Keegan}{0000-0001-8500-4876}{2}, \\
    \authororcid{Wenke Liu}{0000-0002-4014-4477}{2}, 
    \authororcid{Beata Szeitz}{0000-0001-6414-0537}{2}, 
    \authororcid{David Fenyo}{0000-0001-5049-3825}{2},
    \authororcid{Cl\'{a}udio T. Silva}{0000-0003-2452-2295}{1},
    \authororcid{Juliana Freire}{0000-0003-3915-7075}{1}
}

\authorfooter{
\textsuperscript{*}
Contributed equally to this work. \textsuperscript{1}NYU Tandon School of Engineering.
\textsuperscript{2}NYU Grossman School of Medicine.}
\abstract{%
    Biomedical data harmonization is essential for enabling exploratory analyses and meta-studies, but the process of schema matching—identifying semantic correspondences between elements of disparate datasets (schemas)—remains a labor-intensive and error-prone task. Even state-of-the-art automated methods often yield low accuracy when applied to biomedical schemas due to the large number of attributes and nuanced semantic differences between them. We present BDIViz, a novel visual analytics system designed to streamline the schema matching process for biomedical data. Through formative studies with domain experts, we identified key requirements for an effective solution and developed interactive visualization techniques that address both scalability challenges and semantic ambiguity. BDIViz employs an ensemble approach that combines multiple matching methods with LLM-based validation, summarizes matches through interactive heatmaps, and provides coordinated views that enable users to quickly compare attributes and their values. Our method-agnostic design allows the system to integrate various schema matching algorithms and adapt to application-specific needs. Through two biomedical case studies and a within-subject user study with  domain experts, we demonstrate that BDIViz significantly improves matching accuracy while reducing cognitive load and curation time compared to baseline approaches.

}

\keywords{Schema matching, Biomedical data harmonization, Data visualization, User-in-the-loop, LLM-based schema matching}
\teaser{
  \centering
  \includegraphics[alt={Biomedical data harmonization requires experts to manually match attributes across disparate datasets and schemas (1A-C). BDIViz(Biomedical Data Integration Visualization) streamlines this process through interactive exploration and validation of matches using visualization and LLM-based support. BDIViz leverages (2) automated methods for schema matching to (3A-3B) derive candidate matches, which are  summarized in (6) an interactive heatmap. As users investigate and validate matches, the system employs a data-driven approach to empower them to make informed decisions by providing the ability to (4A) compare attribute values, (4C) explore similarity thresholds, and (4B) adjust matcher weights. (5)~The LLM-based agent assists in match exploration and validation.}, width=.98\linewidth, height=0.32\textheight, alt={Overview of BDIViz system interface.}, trim=10pt 15pt 10pt 0pt, clip]{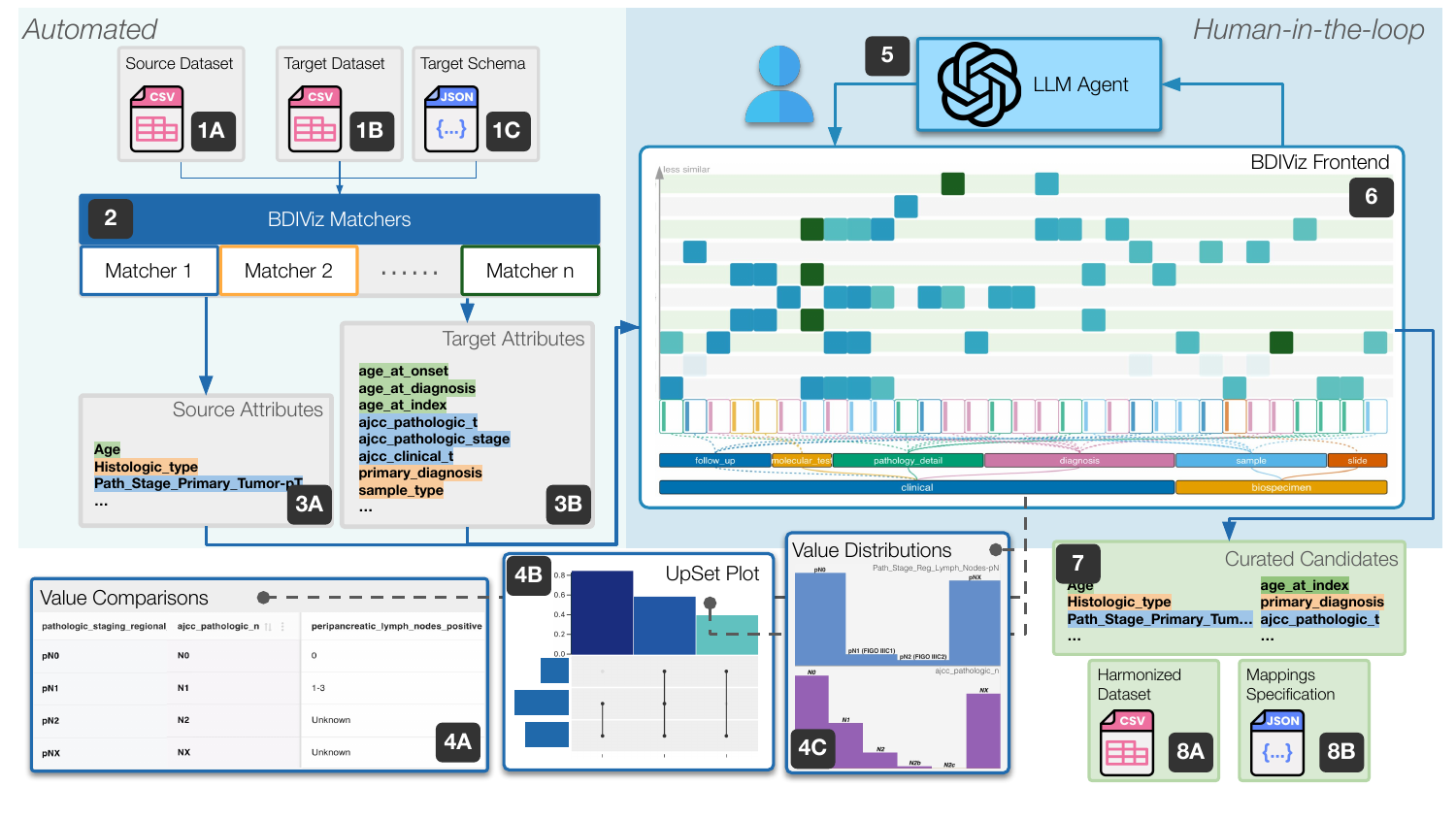}
  \caption{% 
  Biomedical data harmonization requires experts to manually match attributes across disparate datasets and schemas (1A-C).  \toolname (\textit{\textbf{B}iomedical \textbf{D}ata \textbf{I}ntegration \textbf{Vis}ualization}) streamlines this process through interactive exploration and validation of matches using visualization and LLM-based support. \toolname~leverages (2) automated methods for schema matching to (3A-3B) derive candidate matches, which are  summarized in (6) an interactive heatmap. As users investigate and validate matches, the system employs a data-driven approach to empower them to make informed decisions by providing the ability to (4A) compare attribute values, (4C) explore similarity thresholds, and (4B) adjust matcher weights. (5)~The LLM-based agent assists in match exploration and validation
  by deriving explanations. (7) The curated candidates are used to derive the (8A) harmonized dataset and (8B) mapping specification.}
  \label{fig:teaser}

  }
  
\newcommand{\toolname}{\textit{BDIViz}\xspace}

\newcommand{\hide}[1]{}

\newcommand{\myparagraph}[1]{\noindent \textbf{#1.}}

\graphicspath{{figs/}{figures/}{pictures/}{images/}{./}} % where to search for the images

\usepackage{tabu}                      % only used for the table example
\usepackage{booktabs}                  % only used for the table example
\usepackage{lipsum}                    % used to generate placeholder text
\usepackage{mwe}                       % used to generate placeholder figures

\usepackage{mathptmx}                  % use matching math font

\begin{document}
%%%%%%%%%%%%%%%%%%%%%%%%%%%%%%%%%%%%%%%%%%%%%%%%%%%%%%%%%%%%%%%%
%%%%%%%%%%%%%%%%%%%%%% START OF THE PAPER %%%%%%%%%%%%%%%%%%%%%%
%%%%%%%%%%%%%%%%%%%%%%%%%%%%%%%%%%%%%%%%%%%%%%%%%%%%%%%%%%%%%%%%


\firstsection{Introduction}
\maketitle

Schema matching is a critical step in data harmonization (integration) required to identify semantic correspondences between attributes of disparate datasets~\cite{SchemaMatching}. In biomedical research, where datasets are increasingly spread across repositories~\cite{pdc, gdc, pubmed}, harmonization is essential to enable analyses and cross-study comparisons necessary for scientific discovery~\cite{sweeney2023case,grossman2023ten,thangudu2024nci,wang2024nci,stark2024call}. Despite investments to promote data sharing and adherence to the FAIR principles~\cite{wilkinson2016fair},
%to ensure data are findable, accessible, interoperable, and reusable, 
harmonization remains labor-intensive and error-prone%
%, and costly
~\cite{grossman2023ten,sweeney2023case,thangudu2024nci,stark2024call}. %\footnote{https://proteomic.datacommons.cancer.gov/pdc/}
While many \textit{automated schema matching approaches} exist~\cite{freire@deb2025, goguen2005, SchemaMatchingNMappingChpt2, SANTOS, chorus@vldb2024, cong2023pylon, re@vldb2022}, no single method consistently performs well across datasets~\cite{koutras2021valentine, liu2024magnetocombiningsmalllarge}. Users must therefore manually validate matches, eliminate false positives and negatives~\cite{harmonizationPrime2024}.
%, and resolve missed correspondences~\cite{harmonizationPrime2024}. 
This challenge is particularly acute in biomedical schemas, where even the best-performing methods struggle due to scale and ambiguity~\cite{liu2024magnetocombiningsmalllarge}. 

\pagebreak
Standard biomedical schemas, such as the Genomic Data Commons (GDC) with over 700 attributes~\cite{gdc}, including patient demographics, diagnosis, pathology details, family history, are commonly used for data sharing and integration in cancer research. Researchers often map their datasets to these schemas for sharing in data commons and often use them as the target for integrating disparate datasets~\cite{grossman2023ten}. 
For example, in a pan-cancer analysis, \revision{Li et al. integrated datasets from 10 studies of different cancer types~\cite{li2023proteogenomic}, containing a total of 569 attributes, by first mapping them to the GDC schema, a process that took several months}. 
%
%Despite using automated matchers to narrow down candidates, over 383,000 potential matches required manual verification. 
In addition to the very large number of potential candidate matches (over 383,000), 
the task was further complicated by nuanced semantic differences between attributes, e.g., \texttt{age\_at\_diagnosis} and \texttt{age\_at\_index} (\figref{\ref{fig:teaser}}{3A,3B}), thereby making accurate match curation cognitively demanding even for domain experts.

Several \textit{visualization-based tools} have been proposed to assist in general schema matching~\cite{COMA, CLIO, AMC, SystemER}. However the node-link visualization paradigm they used, where source and target attributes are represented as nodes and matches as links, does not scale well for large biomedical schemas and lacks domain-specific semantics needed to resolve terminology ambiguities.  

% but they exhibit significant limitations for biomedical applications. This is because the node-link visualization paradigm used by these tools (where source and target attributes are represented as nodes and suggested matches as links), fail to scale effectively for large schemas. Additionally, these tools do not incorporate domain-specific semantic information necessary for resolving ambiguities in biomedical terminology.

As a result, despite advances in \textit{automated methods} and \textit{visual interfaces}, the identification and validation of matches remains time-consuming and error-prone. Harmonization errors can have serious consequences: they compromise analytical integrity, distort scientific and clinical conclusions, impede cross-study validations, and ultimately hinder research progress. This challenge reveals a critical gap in the design of tools for biomedical schema matching.

\vspace{3 pt}
\myparagraph{User-in-the-Loop AI-Powered Visual Schema Matching}
We propose a novel visual analytics tool -- \toolname -- that integrates automated matching methods, interactive visualization, and LLM-based validation and explanation to address the challenges in biomedical schema matching. To develop an effective solution, we employed a user-centered design approach. We first conducted a formative study with five biomedical experts, identifying key design requirements through interviews and workflow observations (Section~\ref{sec:formative-study}). Then refined \toolname via multiple co-design sessions with three biomedical researchers, incorporating iterative feedback on usability and functionality (Section~\ref{sec:co-design}). 

% This user-centered design approach ensured that our solution addressed the cognitive and practical challenges encountered in biomedical schema matching tasks, while supporting the complex decision-making processes required for accurate data harmonization.

% Scalability --> summarize matches using a heatmap  
\toolname addresses the \emph{scalability challenges} through \emph{novel interactive visualization} techniques. 
It employs an interactive heatmap (\figref{\ref{fig:teaser}}{6}) to summarize matches across large schemas, allowing users to intuitively explore similarities, filter, and refine matches efficiently, and then accept, or reject candidates.
The interface organizes attributes into a tree-like structure, grouping similar attributes for efficient navigation of large schemas.
\toolname \emph{guides users in assessing nuanced semantic differences} through \emph{interpretability and feedback}.
The system provides attribute comparison visualizations (\figref{\ref{fig:teaser}}{4A,4C}), relevant attribute information, and LLM-based match explanations (\figref{\ref{fig:UI}}{3}) to help users disambiguate similar attributes. 
Last but not least, \toolname \emph{captures provenance} by recording user interactions to ensure  transparency,  traceability, and reproducibility of the curation process.
This approach transforms the manual, error-prone process into a guided workflow combining human expertise and computational methods, addressing the needs identified from our formative studies.

% No one-size-fits-all --> combine different matching methods
An important innovation of our approach is its method-agnostic design allowing the integration of various schema matching algorithms.
This flexibility ensures that \toolname can readily incorporate new developments in schema matching methods~\cite{freire@deb2025}. % \revision{as well as domain-specific methods}.
Recognizing that no single method performs consistently well across diverse schemas, \toolname employs an ensemble approach that combines multiple matching algorithms and uses LLMs as an additional matcher.
This multi-faceted strategy offers important benefits: it identifies high-confidence matches (where multiple matchers agree),  directs expert attention to uncertain cases that require human judgment, leverages the broad knowledge encoded in LLMs, and as we show in Section~\ref{sec:technical-evaluation}, results in matching accuracies that are higher than what is attained by individual matchers used in isolation. 
%Through, and takes advantage of the broad knowledge encoded in LLMs to aid in match curation. 
%
Users can inspect matcher influence, adjust weights, and refine rankings (\figref{\ref{fig:teaser}}{4B}), ensuring transparency and control throughout the curation process.

%We carried out a quantitative performance evaluation that confirms the effectiveness differen \jf{xxxx}
We evaluated \toolname with a within-subject user study with twelve expert participants, comparing its performance against a baseline workflow. Participants completed two schema matching tasks in a counter-balanced order.
% We measured their completion time, matching accuracy, system usability score (SUS), and cognitive workload index (using NASA TLX)~\cite{nasaTLX}.  
NASA TLX~\cite{nasaTLX}, SUS, and performance measures results showed that \toolname reduced task completion time by over 14\% while significantly improving accuracy (by 50\%). We also show that \toolname generalizes and addresses large scale complex schema matching tasks for real-world use cases in cancer research. 
%, such as the Biomedical Data Fabric (BDF) program that supports data sharing and re-use. 
We  carried out a quantitative evaluation of different system components and present case studies that illustrate 
the system's effectiveness across diverse real-world biomedical schema matching tasks.
% \jf{Mention that the system was developed in the context of the BDF program, which aims to build a biomedical data fabric -- infrastructure to support data sharing and re-use}

\myparagraph{Contributions}
Our main contributions can be summarized as follows:
\begin{itemize}[leftmargin=*, itemsep=-2pt]

\item A user-centered design approach, using a formative study and iterative co-design sessions with domain experts to identify key domain-specific design requirements for schema matching (Section~\ref{sec:study}).

\item \toolname, \revision{an open-source} visual analytics system that tackles scalability, semantic ambiguity, and usability challenges through interactive heatmaps \revision{and filtering}, comparative attribute visualizations, and LLM-derived explanations (Section~\ref{sec:system})~\cite{bdiviz-github}.

\item A method-agnostic ensemble that integrates multiple matchers with LLM-based validation to identify high-confidence matches and direct expert attention to uncertain cases.

\item Evaluation of the system's performance -- a within-subject expert user study (n=12) and two biomedical case studies (Section~\ref{sec:technical-evaluation},~\ref{sec:user-evaluation},~\ref{sec:case-study}). 
\end{itemize}

\section{Background and Related Work}

%\claudio{I suggested to Juliana that we have a figure that summarizes the problem that we are trying to solve: schema matching for biomedical applications}
%
%\eden{We did show it in teaser 3A, 3B, and 7. Shall we add a seperate one?}
%
%\dt{I have rarely seen figures for background and RW, might be better to just call out the teaser fig here}

Schema matching is a widely studied problem and an important step required to integrate data from multiple sources. 
% , with a variety of methods applicable in different settings.
%Below,  we briefly summarize the related work  and discuss the limitations of existing schema matching systems with dedicated GUIs.
%We also discuss 

%\subsection{Schema Matching Methods}
%\jf{I have reworded this paragraph}
\myparagraph{Schema Matching Methods} 
Schema matching encompasses methods for determining the correspondence between elements from different schemas (tables).
% JF: this sounds repetitive
% In this work, we focus on tabular data where schema matching consists of finding relationships between attributes.
Typically, \revision{a schema matching method receives a source and a target table as input}, computes a set of similarity metrics for attribute pairs, and outputs a subset of pairs representing potential matches. Earlier works in the field relied mainly on syntactic similarity measures between attribute names and instances~\cite{rahm2001survey, zhang2011automatic}. Recent research has employed techniques that capture semantic similarity between attributes. 
% , based on representations produced by modern embedding models.
Some methods employ pre-trained models, including: \emph{i}) pre-trained word and character embedding models~\cite{fernandez2018seeping, cappuzzo2020creating}, \emph{ii}) pre-trained or fine-tuned transformer-based models \cite{zhang2023schema, inSituSchMatcICDE2024, liu2024enhancing}, and \emph{iii}) (self-)supervised deep learning models \cite{zhang2021smat, koutras2024omnimatch}. The emergence of LLMs has led to the development of approaches that utilize prompts to obtain potential matches between pairs of tables \cite{li2023table, parciak2024schema}; however, such methods are not effective or even applicable when schemas are large, as in the cases we study in this work, due to the cost and context limits of LLMs. More recently, Liu et al. proposed a new method that attemps to address this limitation by combining pre-trained models and LLMs~\cite{liu2024magnetocombiningsmalllarge}.
%\jf{\cite{liu2024magneto} combines pre-trained models and LLMs and was shown to be outperform prior state-of-the art methods for biomedical schemas -- we have integrated this into our prototype}

Despite the multitude of schema matching techniques in the literature, experimental evaluations have shown that there is no method that consistently performs well for different data and schemas~\cite{koutras2021valentine, liu2024magnetocombiningsmalllarge}. Therefore, we have designed \toolname to combine the results of several schema matching methods and provide vizualizations that intuitively summarize the outputs of individual matchers. % and enable users to control how the methods are combined.
Furthermore, %since no automated method is foolproof, % jf this sounds redundant
we leverage LLMs and the broad knowledge they encode to derive matching explanations and help with the selection of valid matches. 
%This was inspired by the strategy introduced in~\cite{liu2024magneto} which used LLMs to successfully prune and rank matches derived by pre-trained models. 

%\subsection{Schema Matching GUIs} 
%\myparagraph{Schema Matching GUIs}
\myparagraph{User Interfaces for Schema Matching} 
Some schema matching systems provide GUIs that
% support visual exploration and match verification:
enable users to visually inspect matches %between elements of a source and a target schema,
derived by the underlying matching method. The COMA 3.0 GUI \revision{displays matches as a bipartite graph and} allows users to review and edit automatically generated matches by manually adding and removing edges between attributes~\cite{COMA, COMA++}.
BizTalk Mapper introduced incremental matching through its GUI, allowing users to inspect candidate matches for each source attribute one at a time~\cite{IncrementalSchemaMatching}.
% in the source schema in isolation.
AMC provides more flexibility: users can execute and debug matching processes by moving forward and backward, setting breakpoints, and analyzing intermediate matching results through visualizations and 3D histograms to check similarity scores~\cite{AMC}. Harmony presented a schema matching system that is enhanced by a GUI for filtering results~\cite{mork2008harmony, seligman2010openii}. Unlike the aforementioned systems, Valentine uses a GUI to enable the execution of several schema matching methods and allows
%among any given number of tabular datasets. The 
users to inspect lists of attribute pairs ranked by the similarity score that each individual method outputs~\cite{koutras2021valentinedemo}.

% Nonetheless, existing systems with GUIs have important limitations. First, they employ simplistic visualizations, such as lines that represent candidate matches, which can hinder user inspection and interaction, especially for large source and target schemas. Furthermore, users are typically presented with similarity scores between attribute pairs as the only insights to help them decide on potential matches; consequently, users not familiar with the workings of underlying methods cannot interpret such scores. We also note several of the systems we reviewed rely on results of a single matching method, and many are described in papers but are not available for use. Our proposed open-source tool overcomes these limitations, by \emph{i}) introducing intuitive and insightful visualizations, \emph{ii}) supporting matching of large schemas without overwhelming the users, \emph{iii}) providing insights based on data characteristics, enhanced by LLMs, and \emph{iv}) seamless integration of matching methods.

These schema matching GUIs have important limitations.
\revision{First, most use node-link diagrams or line-based visualizations to represent matches, leading
to visual clutter and making it hard to navigate as the number of schema attributes increases. Ghoniem et al. and Nobre et al. showed that node-link diagrams are especially limited in dense or large-scale scenarios, like in schema matching,  due to cognitive overload and edge overlap~\cite{ghoniem2005readability, nobre2019state}.
Furthermore, users are typically presented with similarity scores between attribute pairs as the only guidance for evaluating potential matches. Kandel et al. emphasized that abstract statistics hinder understanding and decision-making, especially for users without deep context knowledge~\cite{kandel2011research}.
\toolname overcomes these limitations by: \emph{i}) introducing intuitive, a matrix-based visualization to display and summarize matches, \emph{ii}) supporting large-scale schema matching without overwhelming users,
\emph{iii}) providing data- and context-based insights enhanced by LLMs, and \emph{iv}) enabling the integration of multiple matching methods.
}

\revision{
\myparagraph{Scalable Heatmap Visualization}
Schema matching results are naturally represented as a 2D matrix of scores between columns from two database schemas~\cite{COMA}, making heatmaps a suitable visual representation. However, real-world schemas often involve tens to over a hundred columns~\cite{liu2024magnetocombiningsmalllarge}, posing challenges for scalability, grouping, and drill-down analyses.
While conventional heatmaps can scale to wide matrices~\cite{gove2011netvisia, fernandez2017clustergrammer}, they often lack support for column grouping and focused exploration. Focus+Context techniques are commonly used in visual analytics to enable detailed inspection while preserving global context~\cite{isokoski2018useful}. Prior work has proposed expandable heatmap cells to support such interaction~\cite{horak2021responsive}, allowing users to enlarge a cell on demand.
% However, these designs are not tailored to schema matching tasks and do not present underlying schema data within the expanded view.
%
% To address this, we introduce an embedded design
We make use of this design pattern to enable users to drill down into a match: each expanded cell displays a histogram that summarizes the value distributions in the corresponding columns.
For column grouping, prior work often combines heatmaps with hierarchical visualizations such as node-link diagrams or dendrograms~\cite{tiessen2017improved, sun2022single}. While effective for small hierarchies, these methods can become visually cluttered and disconnected from the heatmap when applied to large schemas. To improve visual coherence and scalability, we adopt a space-filling treemap layout~\cite{shneiderman1992tree} to represent column clusters. Unlike node-link diagrams, space-filling treemaps make efficient use of screen space and allow for direct alignment with the heatmap grid~\cite{schulz2011treevis, li2023gotreescape, li2015exploring}. In our system, the space-filling treemap structure is presented adjacent to the heatmap axes, providing users with an integrated view of both column grouping and matching scores. This design supports scalable comparison while maintaining a compact and coherent visual layout.
}

%% TO ADD - -from reviews
% - StratomeX: Visual Analysis of Large-Scale Heterogeneous Genomics Data for Cancer Subtype Characterization https://doi.org/10.1111/j.1467-8659.2012.03110.x
% - Vimo: Visual Analysis of Neuronal Connectivity Motifs https://doi.org/10.1109/tvcg.2023.3327388
% - HiPiler: Visual Exploration of Large Genome Interaction Matrices with Interactive Small Multiples https://doi.org/10.1109/TVCG.2017.2745978
% - Lineage: Visualizing Multivariate Clinical Data in Genealogy Graphs https://doi.org/10.1109/TVCG.2018.2811488
% - Screenit: Visual Analysis of Cellular Screens https://doi.org/10.1109/TVCG.2016.2598587
% - Vials: Visualizing Alternative Splicing of Genes https://doi.org/10.1109/TVCG.2015.2467911
% - Peax: Interactive Visual Pattern Search in Sequential Data Using Unsupervised Deep Representation Learning https://doi.org/10.1111/cgf.13971
% - UpSet: Visualization of Intersecting Sets https://doi.org/10.1109/TVCG.2014.2346248
% - MizBee: A Multiscale Synteny Browser https://doi.org/10.1109/tvcg.2009.167

\revision{
\myparagraph{User Interfaces for Data Wrangling}
Data wrangling presents fundamental challenges for data exploration and  visualization.
It entails many different operations, including data transformation, schema matching, entity resolution, and data cleaning~\cite{kandel2011research}. 
Different systems have been proposed for different tasks. For example, 
Kendel et al. pioneered interactive data wrangling through Wrangler~\cite{kandel2011wrangler}, which allows users to specify data transformations through direct manipulation. Rigel introduced a
declarative approach to data transformation that relies on the visual arrangement of tables -- rows, columns and cells~\cite{chen2023rigel}. 
%\jf{should we add references to visual systems for data cleaning? e.g., open refine? 
OpenRefine allows users to iteratively apply data cleaning operations while interacting with the data in a spreadsheet view~\cite{openrefine}.
%extended this approach to support declarative table mapping.
% These systems focus on within-dataset transformations. In contrast, our work adapts their human-in-the-loop and data-centric principles to the challenge of cross-dataset schema matching in the biomedical domain.
These systems aim to help users carry out a series on individual data transformation steps and do not support automated schema matching. While \toolname also employs a user-in-the-loop approach, it does so to support the exploration and verification of matches derived by automated schema matching methods.
} 

\revision{
\myparagraph{Graph and Relational Data Visualization}
Schema matching can be seen as a graph visualization problem since it
entails the visualization of relationships between two sets of entities (source and target attributes). Ghoniem et al. demonstrated that matrix representations outperform node-link diagrams for dense graphs and relationship analysis tasks~\cite{ghoniem2005readability}. 
Henry and Fekete further showed the benefits of dual-representation systems, combining node-link diagrams and matrices for flexible exploration of social network structures~\cite{henry2006matrixexplorer}.
Nobre et al. surveyed techniques for visualizing multivariate networks, emphasizing the advantages of adjacency matrices for detailed and scalable attribute analysis~\cite{nobre2019state}.
Additionally, Alper et al. provided insights into effective techniques for comparing weighted graphs, highlighting matrix-based visualizations as particularly effective for complex relationship assessments~\cite{alper2013weighted}.
Our approach builds upon these foundational studies, employing a matrix-based representation to effectively handle dense biomedical schema matching scenarios.
}

% \input{sections/03-background}
% \newpage
\section{Formative Study and Co-Design}
\label{sec:study}
% \jf{We have not introduced value mapping: we have two choices -- 1) add this to the intro, 2) discuss this in section 4 when we present the tool, explaining that harmonization requires both finding correspondences between attributes and mapping values into the required standard. For now I commented out the reference to value matching}
% \dt{makes more sense to discuss in Sec 4 with the details}
%
To design a tool that aligns with biomedical researchers' data harmonization workflows,
%—specifically for schema and value matching—
we conducted a formative study with biomedical experts and data scientists. Our goal was to understand:  (1) their existing schema matching processes, and (2) identify key challenges they face, including current workflow inefficiencies and pain points.

% % \vspace{4 pt}
% \input{tables/experts}
% % \vspace{-10 pt}

\myparagraph{Expert Participants} 
We recruited five expert participants from our collaborator network. Among these experts, three are biomedical researchers and data scientists from a major academic medical institution (E1 - E3), one is a principal scientist from a major research organization (E4), and one is a data librarian at a prominent academic medical library (E5). \textit{(Note: a detailed table is provided in the supplementary material).}

% \vspace{-15 pt}

\begin{figure}[h!]
    \centering
    \includegraphics[alt={User-centered design approach. Left: Schema matching requirements (R1-R7) derived from expert interviews, color-coded (red: match discovery, green: value analysis, yellow: filtering). Right: Design considerations developed in iterative co-design sessions with domain experts DC1-DC5. Colored lines between panels indicate the requirements and their corresponding design considerations.}, width=\linewidth]{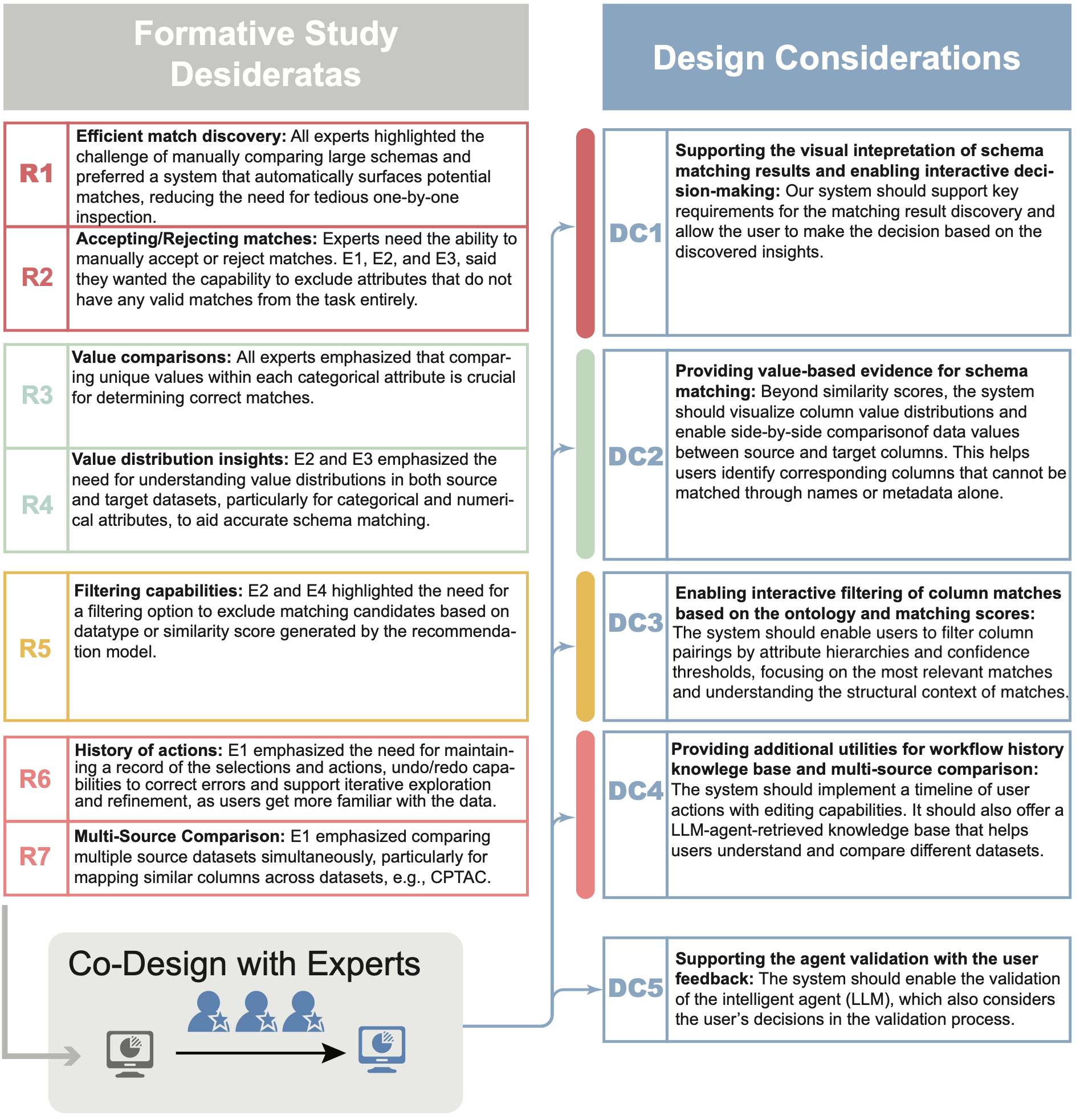}
        \vspace{-.5cm}
    
        \caption{User-centered design approach. Left: Schema matching requirements (R1-R7) derived from expert interviews, color-coded (red: match discovery, green: value analysis, yellow: filtering). Right: Design considerations developed in iterative co-design sessions with domain experts (\dclabel{dc:desiderata}{DC1}-\dclabel{dc:llm}{DC5}). 
          %, \dclabel{dc:values}{DC2}, \dclabel{dc:intuitive}{DC3}, \dclabel{dc:additional}{DC4} and \dclabel{dc:llm}{DC5}). 
    % \guande{cannot remove them because of the hyperref in the later text.} 
    Colored lines between panels indicate the requirements and their corresponding design considerations. 
    % \jf{the text for DC3 mentions ontology, this should be replaced with attrib hierarchy; there are two dots in the end of DC1; }
    }
    \vspace{-.5cm}
    \label{fig:desiderata} 
\end{figure}

\label{sec:formative-study}
\myparagraph{Formative Study}
We then conducted a $\sim$30 minutes formative study with each expert. The study began with the experts first completing a questionnaire on how they approach schema matching, key evaluation factors, and tools used. Through a semi-structured interview, we then collected feedback on their workflow, current challenges, and feature suggestions. Each 30-minute interview was transcribed and analyzed to extract key insights, which we synthesize into a list of prioritized user needs or desiderata for the tool design (Fig.~\ref{fig:desiderata}).  
% \jf{There is a "-" in compar-ing in R3 and in capabili-ties in R6, and in particu-larly in R7; in R6 "E1 pointed to the need
% maintain a record of their actions" fix "the need to maintain"; in R4: "to help accuracy" to help determine correct matches}

% \input{tables/desiderata}

        % \dcanchor{dc:desiderata}
        % \dcanchor{dc:values}
        % \dcanchor{dc:intuitive}
        % \dcanchor{dc:additional}
        % \dcanchor{dc:llm}

\begin{figure*}[t]
\begin{center}

  \includegraphics[alt={The BDIViz interface includes: (1A) a shortcut panel for managing matching candidates, undo/redo, importing datasets, and exporting results as CSV or JSON; (1B) a control panel for filtering candidates; (1C) a timeline graph showing the history of user actions; (2A) an interactive heatmap panel displaying matching candidates with source attributes on the y-axis and target attributes displayed using a space-filling treemap layout on the x-axis;  selected matches (cells) expand to show value distributions; (2B) bottom: an UpSet plot, value comparisons, and parallel coordinates for understanding the source-target relationships; (3A)  agent explanation panel showing LLM validation results with reasoning (e.g., semantic match, shared tokens/values, historical references); (3B) display of target attribute properties, including node name, category, type, descriptions, and value distributions; and (4) a search bar to filter matches using keywords.}, width=.9\textwidth]{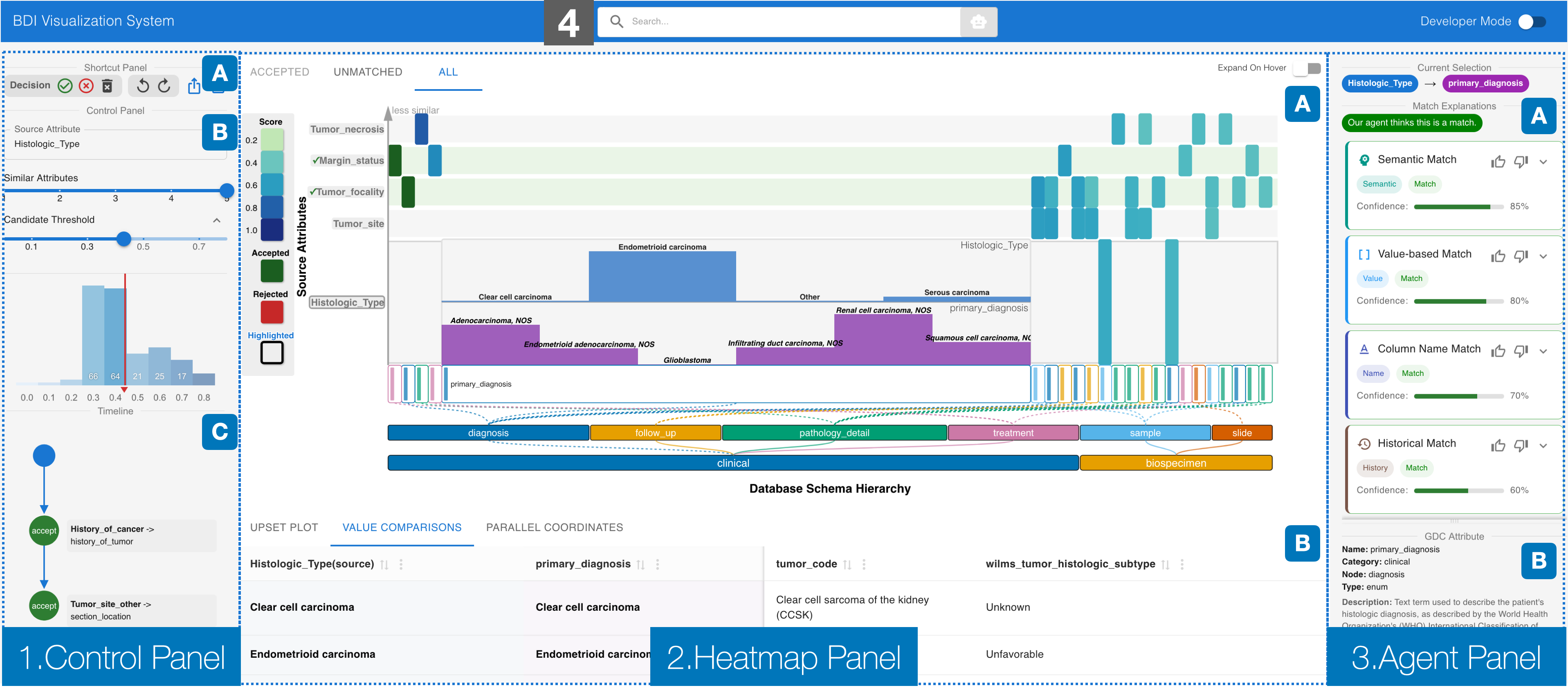}
    \vspace{-.2cm}
    \caption{\revision{The \toolname interface includes: (1A) a shortcut panel for managing matching candidates, undo/redo, importing datasets, and exporting results as CSV or JSON; (1B) a control panel for filtering candidates; (1C) a timeline graph showing the history of user actions; (2A) an interactive heatmap panel displaying matching candidates with source attributes on the y-axis and target attributes displayed using a space-filling treemap layout on the x-axis;  selected matches (cells) expand to show value distributions; (2B) bottom: an UpSet plot, value comparisons, and parallel coordinates for understanding the source-target relationships; (3A)  agent explanation panel showing LLM validation results with reasoning (e.g., semantic match, shared tokens/values, historical references); (3B) display of target attribute properties, including node name, category, type, descriptions, and value distributions; and (4) a search bar to filter matches using keywords.}}\label{fig:UI}
\end{center}
\vspace{-.7cm}
\end{figure*}

Key takeaways 
%for tool design 
include the necessity of visualization features for efficient match discovery using visual representations like heatmaps and histograms, 
direct manipulation of matches (accept/reject), support for iterative data exploration and debugging using value comparisons and value distributions, filtering and prioritization of attributes in large biomedical schemas, and the  recording of actions performed. Based on these insights, we developed the initial prototype of the tool that was further refined iteratively through the following co-design sessions with three experts.

\label{sec:co-design}
\myparagraph{Co-Design with the Experts}
We conducted three one-hour co-design sessions with biomedical researchers to ensure our system's usability and alignment with their needs, following a user-centered design approach. 
\textit{\textbf{First Session:}} We presented an initial prototype
% for schema matching,
using a source dataset from Clark et al. to match the Genomic Data Commons (GDC) schema~\cite{clark2019integrated}. Feedback was gathered on usability, complexity, and areas of confusion, which informed the UI improvements. 
\textit{\textbf{Second Session:}} Based on initial feedback, we refined the heatmap, updated color schemes, introduced a semantic clustered tree structure for labels, and added a preliminary LLM-based agent. We listed 14 UI features and sought input on their importance (for example, critically needed, good to have, could be removed, or needed redesigning), which helped us prioritize design changes.
\textit{\textbf{Third Session:}} Using feedback from the second session, we improved the LLM agent, visual layout, interactivity, and added a history feature. The final session provided further refinements to ensure \toolname met researchers' needs.
Throughout all the co-design sessions, experts shared insights and use cases, guiding design improvements. 
%
% \input{tables/designconsiderations}
% These design considerations are reflected in the final version of our system - \toolname, which we discuss in further depth in the following section. 
The co-design sessions culminated in the design considerations shown in \revision{Fig. \ref{fig:desiderata} and the} \toolname system described next. 

\section{\toolname: AI-Powered Visual Schema Matching}
\label{sec:system}

%We introduce \toolname, a biomedical schema-matching tool that combines interactive visualization for match exploration with an LLM-based agent for validation. In this section, we outline the system’s overview workflow, the user interface, the system's backend processing, and the technical implementation.

% We present \toolname~- a biomedical data integration tool for schema matching that uses interactive visualization to support match exploration and an LLM-based agent to support match validation. In this section, we present the overview of the system using the schema-matching workflow and describe the interface features, backend processing, and technical implementation.

\revision{
\myparagraph{Design Rationale}
Our design adopts a matrix-based heatmap visualization combined with interactive filtering and semantic overlays to address scalability and interpretability challenges in schema matching. Prior studies have shown that matrix-based representations scale more effectively than node-link diagrams, preserving spatial consistency and enabling pattern recognition across large attribute sets~\cite{ghoniem2005readability, holten2006hierarchical, henry2006matrixexplorer, nobre2019state}. To further support user navigation and reduce visual complexity, we incorporate interactive filtering, allowing users to focus on high-confidence matches or subsets of interest — an approach aligned with principles of faceted interaction and details-on-demand~\cite{heer2012interactive, wongsuphasawat2016voyager}.
Additionally, our system supports human–AI collaboration by integrating LLM-powered semantic explanations to help users interpret similarity scores and make informed decisions. This reflects recent HCI research on enhancing transparency and trust in AI-supported interfaces and ensures that non-expert users can meaningfully engage with the matching process~\cite{wang2019human, endert2017state, sacha2017what, hohman2020understanding}.
}

\myparagraph{System Overview}
As Fig.~\ref{fig:UI} illustrates, to initiate a data harmonization task, the user uploads the source and target datasets or a source dataset and a target schema  (\figref{\ref{fig:UI}}{1A}).  \toolname processes these inputs using an ensemble of matchers to generate
candidate matches with associated confidence scores (Section \ref{sec:backend}).
%
%To streamline the process, 
The system first identifies and automatically accepts matches that have high confidence scores and that multiple matchers agree on. This allows users to focus their attention on uncertain candidates that are more challenging to validate. 
Users can inspect the accepted matches and modify them if needed. 
% all pairs not just the remaining...
The candidate pairs
% between source and target columns
are summarized in the interactive heatmap matrix (\figref{\ref{fig:UI}}{2A}), which displays the source attributes as rows and target as columns; cells correspond to matches and their color intensity indicates matching confidence. 

Users can configure the datasets/schemas, matcher parameters as well as modify matching thresholds in the \emph{Control Panel} (Fig.~\ref{fig:UI} (1)). 
The \emph{Interactive Heatmap Panel} (Fig.~\ref{fig:UI} (2A)) provides a summary of the matches and enables users to explore and compare them. By selecting a cells, users can examine a specific attribute pair in detail. 
To facilitate exploration, the target attributes are clustered and displayed as a space-filling treemap layout  %and displayed in \emph{Space-Filling Axis} 
(\figref{\ref{fig:UI}}{2A} bottom).
%\jf{we don't talk about ontologies; and I cannot find Space-Filling Axis -- what is this?}
%The panel is adjacent to \textbf{Space-Filling Axis}, which shows the ontology and column relationships with the space-filling tree (Fig~\ref{fig:UI}-2A). 
When a cell is selected,  details of the match are displayed. Here, the selected cell represents a candidate match
between the source attribute \texttt{Histologic\_Type} and \texttt{primary\_diagnosis} from GDC; the value comparison for these attributes is shown in \figref{\ref{fig:UI}}{2B}. In addition, contextual information is displayed in the \emph{Agent panel} to guide accept/reject decisions: explanations
generated by the LLM-based agent  that take into account semantic similarity, shared tokens, and value overlap (\figref{\ref{fig:UI}}{3A}); and attribute descriptions (when available) -- here, the description of \texttt{primary\_diagnosis} from GDC is shown (\figref{\ref{fig:UI}}{3B}).
\revision{Users can filter matches by confidence thresholds, search using keywords (Fig. 3 (4)), and group similar attributes (Fig. 3 (1B) and Section 4.2) to focus on relevant subsets rather than viewing all possible matches simultaneously. The system also supports pagination when the number of selected source attributes exceeds the available screen space.
}
% The system supports this decision-making process with additional visualizations. The UpSet plot displays the distribution of matcher scores across candidates, while value comparison views offer fine-grained visualization of correspondences between attributes. As users interact with the system, their decisions are recorded in a timeline view, allowing for operation tracking and potential rollback.

% JF removed -- this broke the flow, we can connect to the formative study
% when we discuss the individual components
%Through this iterative workflow, \toolname addresses the requirements for streamlining schema matching uncovered in our formative study (Section~\ref{sec:formative-study}).

%from a tedious manual task into a streamlined process. interactive exploration to achieve accurate biomedical data integration. 
%The space-filling axis for ontology representation and in-cell histograms support value matching assessment, addressing the requirement for additional contextual information identified in our formative study (Fig~\ref{fig:desiderata}).

% \textbf{Candidate Quadrant.} In order to further streamline the matching process (DC1), we introduce the \emph{Candidate Quadrant} component to automatically accept easy matches by default.

%\subsection{\toolname UI Features}\label{ssec:ui}
\subsection{UI Components}\label{ssec:ui}
% We next detail some of the features of our interface that support in-depth data exploration for identifying and validating matches. 

In what follows, we describe the UI components in detail and the design considerations they address (Fig.~\ref{fig:desiderata}).

%\subsubsection{Control Panel}
\myparagraph{Control Panel (\dcref{dc:desiderata}, \dcref{dc:values}, \dcref{dc:intuitive})}
The control panel (\figref{\ref{fig:UI}}{1}) serves two primary functions: configuration of matching parameters and keeping track of user actions.
The configuration view enables users to adjust key parameters that influence the automated matching process. Users can select scoring algorithms, threshold values, and weighting factors for different similarity metrics (\figref{\ref{fig:UI}}{1B}). This allows users to tailor the matching process to their specific domain requirements without requiring in-depth technical knowledge of the underlying algorithms.
The provenance component records user interactions chronologically, allowing users to revisit previous states of the curation process (\figref{\ref{fig:UI}}{1C}). Each interaction is represented as a discrete event on the timeline, with icons indicating the type of action performed (e.g., parameter adjustment, match confirmation, rejection). This approach was chosen in addition to alternative designs—such as a simple list of actions or undo/redo buttons (which are also present in the control panel (\figref{\ref{fig:UI}}{1A})) — as it provides both temporal context and the ability to directly navigate to specific decision points. 

% The Control Panel directly supports the configuration and refinement phases of the co-adaptive workflow by allowing users to adjust system behavior based on their evolving understanding of the data relationships.

%\subsubsection{Interactive Heatmap Panel}
\myparagraph{Interactive Heatmap Panel (\dcref{dc:desiderata})}
The Interactive Heatmap View (\figref{\ref{fig:UI}}{2A}) is the core visualization for exploring matches.
%and addresses the design consideration \dcref{dc:desiderata}. 
To visualize complex  relationships among multiple attributes simultaneously, this view uses a matrix-based layout where color intensity encodes similarity scores, with adjustable palettes for accessibility. These color scales are normalized with appropriate padding to enhance visual discrimination between similarity scores. Combining overview and detail-on-demand, this view helps users 
%efficiently 
identify, explore, and validate potential matches.
%at the value level.
As users interactively explore matches, detailed information that captures different aspects of a match are displayed. 

% The combination of overview and detail-on-demand, the Interactive Heatmap View supports the exploration phase of the schema matching workflow, enabling users to efficiently identify potential matches and investigate them at value-level detail before making decisions.

\begin{figure}[t]
    \centering
    \includegraphics[alt={(1) Interactive Heatmap cells; (2) Expanded view of the value distribution histograms for the source (2A) and target (2B) attributes shown when users click on a cell.}, width=\linewidth,height=0.17\textheight]{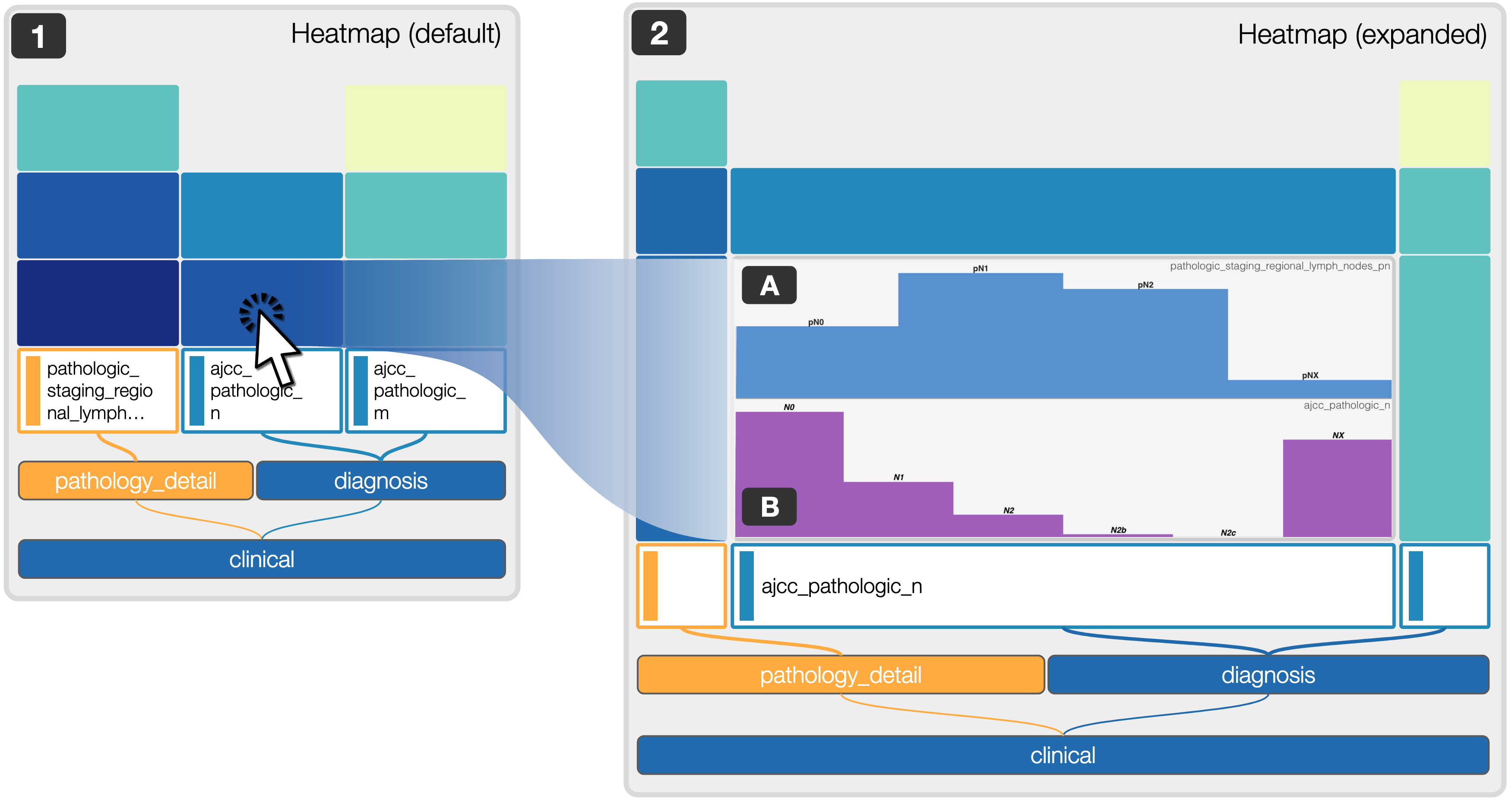}
    \vspace{-.3cm}
    \caption{(1) Interactive Heatmap cells; (2) Expanded view of the value distribution histograms for the source (2A) and target (2B) attributes shown when users click on a cell. 
    % \jf{I thought we had decided not to include this figure -- and we need the space}
    }
        \vspace{-.65cm}
    \label{fig:expansion}
\end{figure}

%\subsubsection{Value Distribution Histogram} 
\myparagraph{Value Distribution Histogram (\dcref{dc:values})} 
%
%The heatmap's interactivity enhances exploration by 
When a user hovers over a cell, it is expanded to display side-by-side histograms with source and target value distributions (\figref{\ref{fig:expansion}}{2A, 2B}). This allows users to go beyond similarity scores and verify semantic alignment by inspecting the actual data values. For example, they can quickly determine if cancer diagnosis codes share similar distributions.

\myparagraph{Target Attribute Hierarchy (\dcref{dc:intuitive}, \dcref{dc:additional})}
The x-axis of the heatmap displays a hierarchical organization of target attributes (\figref{\ref{fig:UI}}{2A}), providing a concise visualization of relationships present in complex biomedical schemas like the Genomic Data Commons (GDC). It employs a three-level hierarchy: supercategories (e.g., "clinical," "biospecimen"), categories within each supercategory (e.g., "diagnosis," "treatment"), and individual columns. The color encoding differentiates categories, while curved connectors represent hierarchical relationships. Our interactive hover functionality allows users to highlight specific levels, super-categories, or individual columns to better understand their relationships. To accommodate schemas of varying sizes, the adaptive layout automatically adjusts spacing, compressing attributes and selectively omitting text labels when necessary. This approach ensures clear, efficient navigation of large schemas containing a large number of attributes while preserving structural context.

\myparagraph{Value Comparison and Mapping (\dcref{dc:values}, \dcref{dc:additional})} 
%
% \dt{Need to add text for this }  NO NEED FOR FIGURE HERE
Users can compare the values of source and target attributes side by side using this component ((\figref{\ref{fig:UI}}{2B})  which presents a one-to-one mapping between the unique values from the source column and the corresponding values from the candidate target column.
% with the highest similarity score.
These mappings can be computed with general similarity \revision{ methods such as RapidFuzz algorithm from PolyFuzz} or specialized libraries~\cite{RepidFuzz, bdi-kit-github}. Note that the leftmost column, labeled "Source Value," remains static, while the column corresponding to the target schema of the selected candidate is dynamically positioned as the second column for easier comparison.

% \subsection{Agent Panel}
% \dt{first - talk about the LLM based explanation generation}
% \dt{second, talk about the 'learning over time' - RAG component - list the three things that the LLM agent learns}

\begin{figure}[t]
    \centering
    \includegraphics[alt={Agent-derived explanations that take different aspect of matches into account.}, width = \linewidth, height=0.23\textheight]{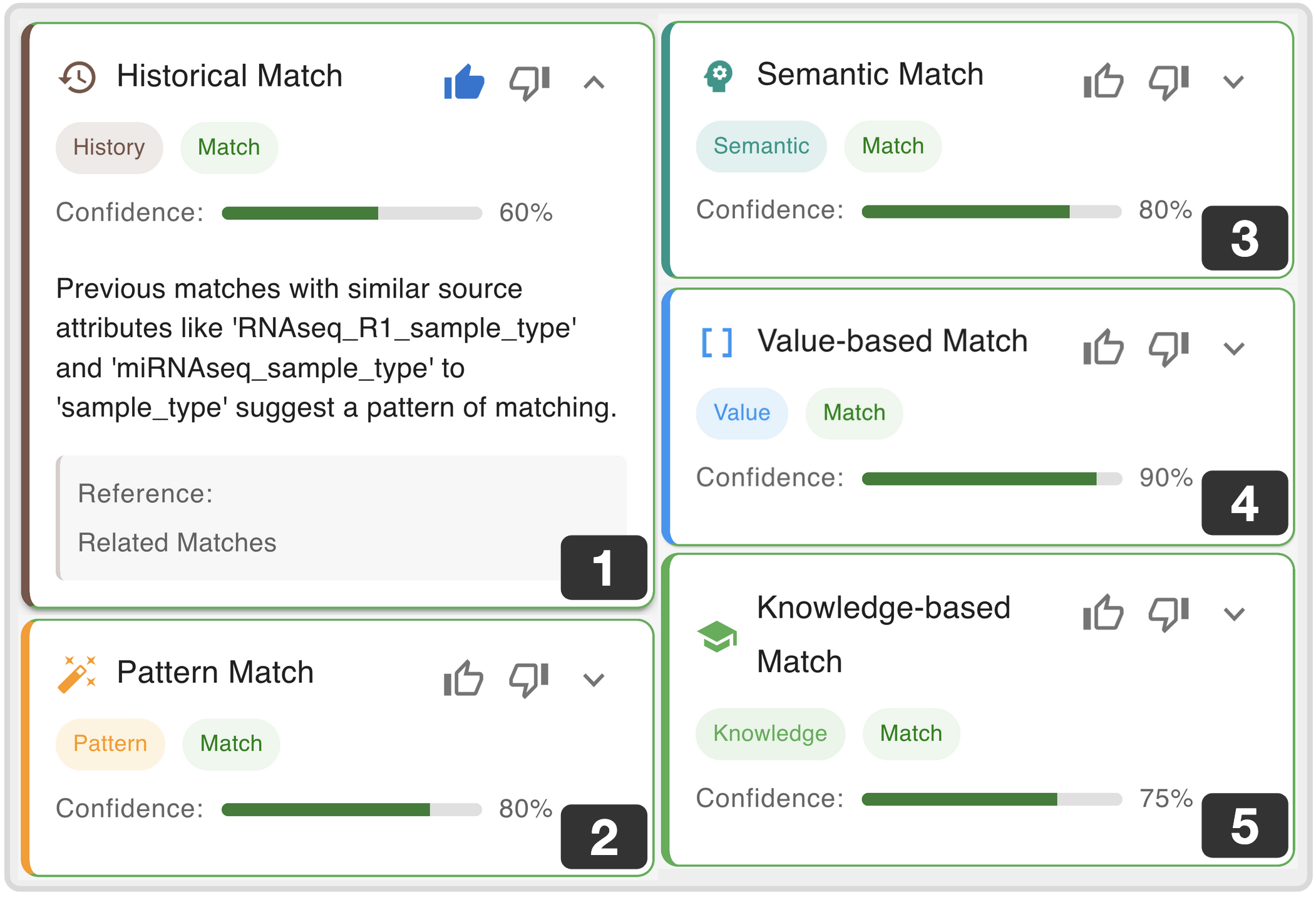}
    \vspace{-.3cm}
    \caption{Agent-derived explanations that take different aspect of matches into account.} %. From (1) to (5) are 5 types of agent explanation cards on validating a matching candidate.}
    \label{fig:explanations}
    \vspace{-.3cm}
\end{figure}

%\subsubsection{Agent Panel}\label{ssec:agent}
\myparagraph{LLM-based Agent Panel (\dcref{dc:llm})} %\label{ssec:agent}
%
% \claudio{I think this might warrant more emphasis.}
%
The Agent Panel serves as the interface between the agent and the user.
%'s domain knowledge using the following features.
\textit{Match Validation with Explanation:}
The panel displays LLM-generated explanations for selected matches using a combination of icons and colors to encode different types of evidence (e.g., semantic match, value distribution similarity, column name similarity). As shown in \figref{\ref{fig:UI}}{3A} and Fig.~\ref{fig:explanations}, each explanation category is presented with a confidence score visualized as a progress bar, helping users gauge the system's certainty. This design allows domain experts to quickly understand the rationale behind suggestions without requiring them to parse complex technical details.
\textit{Knowledge Discovery:} The panel also displays external  knowledge available about the attributes (\figref{\ref{fig:UI}}{3B}), presenting contextual information such as attribute definitions, data types, and value constraints. 

Users can provide feedback on explanations, either confirming the system's reasoning or suggesting corrections, which is incorporated by  the agent. % through rag
The confidence visualization employs a gradient-based encoding derived from the color schemes defined in the system's color utilities. This approach provides an intuitive sense of confidence levels while maintaining visual consistency with the heatmap visualization.
\textit{Agent Learning Over Time:} The LLM agent learns from user interactions over time through a retrieval-augmented generation (RAG) component. This system stores and retrieves: (1) previously validated matches across datasets, (2) domain-specific terminology relationships, and (3) user correction patterns on similar attribute types. By incorporating this historical knowledge, the system progressively adapts its explanations to align with user preferences and domain conventions.

% The Agent Panel directly supports the adaptation phase of the co-adaptive workflow by capturing user feedback and incorporating it into the system's learning process, thereby improving subsequent matching suggestions and building a domain-specific knowledge base over time.

\begin{figure}[h!]
    \centering
    \includegraphics[alt={The UpSet Plot Panel displays (1) matcher weights, (2) the weighted average score for each candidate with the color corresponding to heatmap color palette, and (3) the matchers that produced a given candidate -- shown as dots connected by line.}, height = 1.3 in]{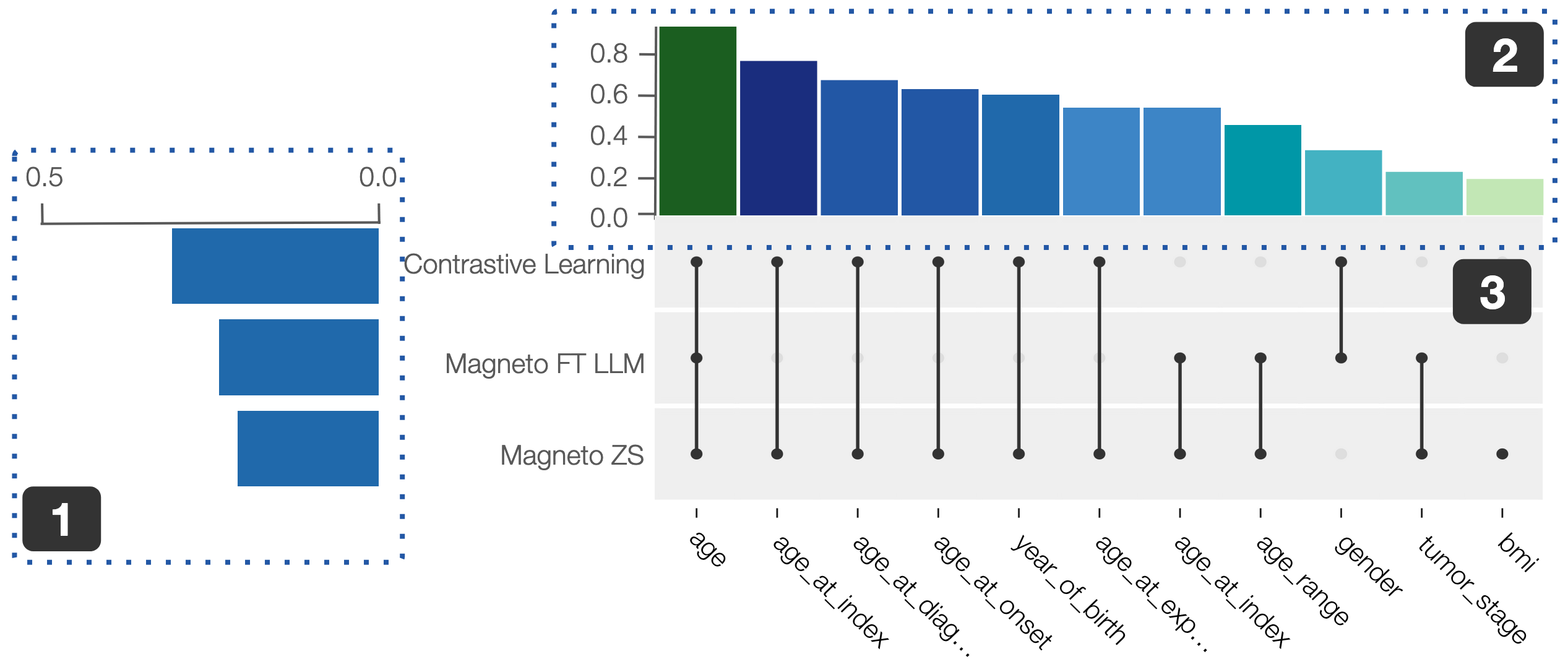}
        \vspace{-.3cm}
    \caption{The UpSet Plot Panel displays (1) matcher weights, (2) the weighted average score for each candidate with the color corresponding to heatmap color palette, and (3) the matchers that produced a given candidate -- shown as dots connected by line.}
    \label{fig:upsetplot}
    \vspace{-.5cm}
\end{figure}

%\subsubsection{UpSet Plot Panel}
\myparagraph{UpSet Plot Panel (\dcref{dc:additional})}
The matching score displayed in the heatmap is computed as the average of an ensemble of matchers. This detailed view of individual matcher contributions is implemented as an UpSet plot visualization, shown in Fig.~\ref{fig:upsetplot}, where each matcher's recommendation is encoded by a dot, allowing users to identify which matchers support a particular match and how these recommendations distribute across candidates~\cite{lex2014upset}. For example, the target attribute \texttt{age} is supported by the three matchers, while \texttt{bmi} is only derived by \texttt{Magneto ZS}. This visualization maintains visual coherence with our interactive heatmap while providing deeper insights into the matching process. 
% JF: removed to save space
%To preserve screen space, the UpSet plot panel is collapsed by default and expands only when users click to open it.
% Upset plot is us

% \newpage
%\subsection{Backend Architecture for Match Generation}
\subsection{Backend and Match Generation}
\label{sec:backend}
% \toolname is built using a robust client-server architecture. The backend, developed with Flask, is responsible for executing automatic schema matchers and managing the LLM agent, while the frontend—constructed with the React Next.js framework—handles data visualization and user interactions. Our visualizations leverage the capabilities of D3 and Observable, ensuring dynamic and interactive displays.
% For schema matching, we integrate multiple libraries, including bdi-kit, Magneto, and Valentine, within our backend. The LLM agent is implemented using LangChain and LangGraph.
% Notably, the entire tool is fully dockerized within a compact Next.js and Flask environment, enabling local deployment and streamlined development.

% \claudio{This section can be substantially shortened. Not clear if there is a need for the workflow backend figure.}

The backend consists of a matching component, the LLM agent that derives explanations, the provenance manager that keeps track of user interactions, and other supporting modules that produce the content required by the UI components. We describe these below.
\myparagraph{Identifying Easy Matches} To prune the space of candidate matches that must be evaluated by matchers and users, \toolname uses a simple heuristic, inspired by what experts do when they manually perform schema matching: it identifies matches that have high similarity for both attribute names and values.  
%alleviate users from having to manually inspect source-target attribute pairs representing easy match cases—those with high attribute name and value similarities—we devise a simple heuristic. 
Specifically, it leverages fuzzy matching scores between attribute names and compute value similarity by averaging the highest fuzzy matching scores for each unique value in the source attribute against the unique values in the target attribute~\cite{RepidFuzz}. Formally, let \( S \) denote the set of unique source values and \( T \) the set of unique target values. If \( f(s,t) \) represents the fuzzy matching score between \( s \) and \( t \), then for each \( s \in S \) we define \( v(s) = \max_{t \in T} f(s,t) \) and compute the value similarity as \( V = \frac{1}{|S|} \sum_{s \in S} v(s) \). We consider a source-target attribute pair as an easy match if both the attribute name similarity and \( V \) exceed predefined thresholds. %we categorize all source-target attribute pairs into four quadrants based on their column name similarity \( C \) and value similarity \( V \). A candidate is deemed an easy match 
%if their attribute name and value similarities are both above a given threshold. 
For these candidates, the matching score is set to 1 and they are subsequently excluded from the candidate set generated by other matchers.

% \label{ssec:ensemble}
\myparagraph{Matcher Ensemble} 
The matching component combines multiple matching methods in an ensemble to derive the match candidates and associated
scores that are displayed in the heatmap matrix.
Our design is matcher-agnostic, making it possible to integrate new schema matching methods as they become available. \revision{Users can either upload a custom matcher as a Python code snippet through the interface or add a matcher script directly to the codebase in a forked repository.} The current prototype 
uses the contrastive-learning method from bdi-kit~\cite{liu2024enhancing}, Magneto~\cite{liu2024magnetocombiningsmalllarge}, and the Jaccard distance matcher from Valentine~\cite{koutras2021valentine}. Matchers are initially assigned equal weights; the weights are automatically adjusted based on their performance, and users can also manually adjust them, e.g., if a given matcher is producing better results.
% JF: if we want to mention that they are automatically updated, we need to explain how and why they are automatically updated
%can be adjusted by the user if desired and will be automatically updated with user operations.

%\myparagraph{Matcher Weight Updater} 
To automatically update matcher weights based on user interactions, we employ the following algorithm. Let candidate \( i \) be associated with a normalized score \( s_i \in [0,1] \) and a rank \( r_i \) among the generated matching candidates, and assume that candidate \( i \) is recommended by a set of matchers \( \mathcal{M}_i \). When a candidate is accepted, the weight \( w_m \) of each matcher \( m \in \mathcal{M}_i \) is updated as \( w_m \leftarrow w_m + \alpha \cdot s_i \cdot \frac{1}{r_i} \); conversely, when a candidate is rejected, the weight is updated as \( w_m \leftarrow w_m - \beta \cdot s_i \cdot \frac{1}{r_i} \), where \( \alpha \) and \( \beta \) are the learning rate parameters for weight increase and decrease, respectively. This mechanism rewards high-quality matchers with strong scores and rankings while penalizing those linked to incorrect recommendations.

\myparagraph{Source Attribute Clustering} To facilitate exploration, the backend leverages embedding techniques such as sentence transformers to cluster source columns based on semantic similarity~\cite{reimers2019sentence}. 
%Additionally, it hosts the LLM agent, responsible for generating matching explanations, and maintains a record of user interactions and historical decisions. Finally, 
%Specialized modules for comparing attribute value distributions are incorporated to enhance the precision and reliability of schema matching.}For source attributes (i.e., columns from the input CSV), we extract feature representations using a Sentence Transformer. 
These embeddings make it possible to apply K-Nearest Neighbors (KNN) clustering algorithm based on cosine similarity to group semantically similar attributes. This clustering acts as a filter, assisting users in exploring related source attributes. 
%Target attributes are not clustered since their ontologies are typically directly extractable from schemas such as PDC and GDC.

%\myparagraph{Value Mappings and Value Bins} For enhanced visualization, \toolname derives both value mappings and value distributions. The value mappings identifies the most similar source-target value pairs for each unique source value in categorical attributes, whereas value bins capture the distribution of unique values for categorical attributes and provide binned counts for numerical attributes. 
%\jf{I am not sure this is necessary, in particular, since we already describe some of this in the UI section, and here there are no technical details}

% Conversely, when a candidate is rejected, the weights of the corresponding matchers are decreased as follows:
% \[
% w_m \leftarrow w_m - \beta \cdot s_i \cdot \frac{1}{r_i},
% \]
% with \( \beta \) representing the rejection learning rate.

% This update mechanism ensures that matchers contributing higher-quality recommendations—those with higher normalized scores and superior rankings—are rewarded, while matchers associated with incorrect recommendations are penalized accordingly.

% \begin{figure}
%     \centering
%     \includegraphics[width=\linewidth]{figs/workflow.png}
%     \caption{\dt{this figure should probably be part of the teaser figure and not a separate one here}}
%     \label{fig:workflow}
% \end{figure}

\myparagraph{LLM-Derived Explanations} 
\toolname leverages a carefully engineered prompt to harness LLM reasoning for schema matching validation.
This prompt integrates contextual constraints, structured data, and historical matching information—managed via a RAG memory keyed by composite identifiers (e.g., “source\_attribute::target\_attribute”). The LLM is then instructed to generate up to four chain-of-thought explanations for each candidate match~\cite{wei2023chainofthoughtpromptingelicitsreasoning}, with each explanation providing a true/false flag, rationale classification (e.g., semantic, name, token, value, pattern, history, knowledge, or other), detailed reasoning, references, and a confidence score. These explanations are then synthesized into a final true/false decision regarding the match's validity.

\subsection{Implementation}

% \claudio{This needs revisions, and we should include a link to the github.}
%\jf{we have no references to the tools we use in the implementation}
\toolname is an \revision{open-source system built using a client-server architecture~\cite{bdiviz-github}}.
%\footnote{GitHub Repo: \revision{https://github.com/VIDA-NYU/bdi-viz-react}} 
%
The backend, developed with Flask, is responsible for executing automatic schema matchers and managing the LLM agent, while the frontend—constructed with the React Next.js framework—handles data visualization and user interactions. Our visualizations leverage the capabilities of D3 and Observable, ensuring dynamic and interactive displays.
For schema matching, we integrate multiple libraries, including bdi-kit~\cite{bdi-kit-github}, Magneto~\cite{liu2024magnetocombiningsmalllarge}, and Valentine~\cite{koutras2021valentine}, within our backend. The LLM agent is implemented using LangChain and LangGraph.
The tool is fully dockerized within a compact Next.js and Flask environment, enabling local deployment and streamlined development.
% \input{sections/07-technical-evaluation}
% \input{sections/05b_llmtab}
% \newpage
\section{Quantitative Performance Evaluation}
\label{sec:technical-evaluation}

%\jf{It is strange to me to call this section "technical" evaluation. Isn't the user evaluation also technical? Would  "Benchmark Evaluation" or "Quantitative Performance Evaluation" be better?}
%
% Also, I think it might be better to do as we had discussed on Friday: have one section called Evaluation with two subsections: "User Evaluation", and then "Quantitative Performance Evaluation"}

% \eden{I have updated the namings accordlingly. However, since there are too many sections and subsections in User Study, I don't think it is a good idea to put it as a subsection. @Dishita, @Guan-de: What do you think?}

% \dt{it is common to have two two evaluation sections in HCI papers - one for technical evaluation and the other being user study}
% \begin{table}[ht]
% \centering
% \scriptsize
% \begin{tabular}{@{}lcccc@{}}
% \toprule
% Dataset & Accuracy & GT Count & GT Covered & GT Coverage \\ \midrule
% GDC Benchmarks & 0.91     & 16.4     & 5.6      & 0.35      \\ \bottomrule
% \end{tabular}
% \caption{Average performance of the Candidate Quadrants Component on the GDC benchmarks. GT Coverage is computed as the ratio of GT Count to GT Covered.}
% \label{tab:avg-performance}
% \end{table}

%Prior to our user study, we showcase the 
To evaluate \toolname,
%of our technical solutions and components, 
we used a benchmark that contains ground truth matches~\cite{liu2024magnetocombiningsmalllarge}, curated by biomedical \revision{experts}, from 10 datasets to the GDC schema. In the following, we discuss results on the effectiveness of \emph{i}) our approach to identify easy matches, \emph{ii}) the matcher ensemble, and \emph{iii}) the match predictions provided by the LLM-agent.

\para{Identifying Easy Matches} To asses the effectiveness of our heuristic to identify easy matches, we compute the overlap between the easy matches and the ground-truth data.
% filtering out easy matches, we set the attribute name similarity threshold to $0.7$ and the value similarity threshold to $0.4$. Then, based on the similarity scores we obtain the amount of correct attribute matches captured using our filtering strategy. 
We observe that $91\%$ of the attribute pairs classified as easy matches annotated as matches are found in the ground truth; the remaining 9\% correspond to cases where a source attribute is \emph{similar} to multiple target attributes.
%describing different kinds of counts. 
%\jf{this is not clear; we should provide concrete examples -- just saying "diff types of counts" is not sufficient for a reader to understand; are these numerical attributes for which there are no values associated with the target attribute?}
%\jf{I also do not understand the sentence below: if a match was identified as easy, it is not processed by the matchers; how is it that multiple matches are derived for it? } \ck{There are cases where the source attribute is matched automatically to more than 1 target attributes, due to high name and value similarity. In these cases the user will have to inspect}
In such cases, the users will have to further inspect them. 
We also see that the correct matches that our 
%simple 
heuristics capture correspond to $35\%$ of all the ground truth pairs, which shows that identifying these easy matches can considerably decrease the user workload.

\para{Matcher Ensemble} We compare the effectiveness of each individual matcher against the ensemble method we employ in \toolname. To do so, for each of the 10 datasets mapped to GDC,  we examine the top-$k$ ($k\in\{10, 20, 40\}$) attribute pairs that each matching method and the ensemble output, with the k-values reflecting pool sizes that domain experts evaluate in practices. Table \ref{tab:matcher-performance} shows the average precision scores across datasets. We see that the ensemble method has significantly higher accuracy than the individual matchers -- it is able to identify  $93\%$ of the ground truth matches in the top-$10$ attribute pairs, while the precision of the individual matchers vary between $70.7\%$ and $81.5\%$. 
%Therefore, matching results based on ensemble scores enable 
%By combining multiple matchers, \toolname to provide the users with a more accurate starting ground, guaranteeing better results than relying solely on individual matchers. 
% by showing their average

\begin{table}[ht]
\centering
\scriptsize
\setlength{\tabcolsep}{4pt}
\begin{tabular}{@{}lccc@{}}
\toprule
Matcher& $k=10$& $k=20$ & $k=40$  \\ \midrule
\toolname Ensemble & \textbf{0.938} & \textbf{0.954} & \textbf{0.965}  \\
ct\_learning & 0.780 & 0.841 & 0.904 \\
magneto\_zs & 0.707 & 0.830 & 0.873 \\
magneto\_ft & 0.815 & 0.869 & 0.911 \\ \midrule
\end{tabular}
\vspace{-.3cm}
\caption{\toolname matcher ensemble average Precision$@k$ score comparison against individual matchers, across 10 different dataset to GDC matching tasks.}
\label{tab:matcher-performance}
\vspace{-.5cm}
\end{table}
\para{LLM-agent} We evaluated the LLM-agent's performance using different OpenAI LLM models (o3-mini, gpt-4o and gpt-4). First, we assessed how well the agent  captures ground truth matches: it incorrectly classifies as non-matches at most $25\%$ of the attribute pairs annotated as matches in the ground truth.
We then evaluated the agent's ability to identify non-matches by testing it on the top-$200$ non-matching attribute pairs with the highest ensemble matching score (hard non-match cases). The results show that the LLM-agent can incorrectly classify up to $38\%$ of them as matches. Similar to the state-of-the art schema matching methods, LLMs are not foolproof, and regardless the choice of model, they make mistakes for challenging schema matching tasks. Nonetheless, they also derive a large number of correct matches and useful explanations. This supports our design decision to combine LLMs with multiple matching methods.

\section{User Study and Evaluation}
\label{sec:user-evaluation}

We conducted a user study to evaluate the effectiveness of \toolname for schema matching tasks. This section details the study design, participant information, procedure, and results.

% \section{User Study}
% Schema matching is a complex task that requires domain knowledge, analytical thinking, and careful attention to detail. To understand how \toolname{} supports users in this challenging process, we conducted a user study involving \dt{12} participants. This study aimed to assess whether the interactive visualization approach of \toolname{} offers tangible benefits in terms of efficiency, accuracy, and user experience compared to traditional methods currently used by biomedical researchers.

\vspace{3 pt}

\subsection{Study Participants}
% To ensure our evaluation reflected the perspectives of the intended user population, we carefully recruited participants with relevant expertise. 
We recruited \emph{12 expert participants} (6F, 5M, 1 nb/ prefer not to say) from departments related to biomedical data science and through snowball sampling. All participants had an average 2 years of experience working with biomedical datasets and performing schema matching or data harmonization tasks, making them well-qualified to evaluate \toolname in our study. Participants' ages ranged from 21 to 27 (M = 24.58, SD = 1.67), and the group covered a diverse range of early to mid-career biomedical data professionals.

% \subsection{Apparatus and Materials}

\subsection{Study Design}
We conducted a within-subjects user study to compare the experience of using \toolname versus traditional schema matching tools for biomedical data integration. Each participant completed two schema matching tasks under the following \emph{conditions:} 
1) using \toolname~as the test condition; 2) using their current tools (e.g., Excel, Python, or R) as the baseline condition, where the same backend-generated matching results were provided in JSON/CSV formats.
We chose different but comparable datasets for the two study tasks to ensure similar complexity while avoiding learning effects. To control for order effects, we counterbalanced the sequence of conditions across participants. The entire study session lasted  $\sim$60 minutes, including the pre-study consent process, tutorial, study tasks, post-study questionnaires, and interview.

\vspace{3 pt}

\noindent \textbf{Study Tasks:} For the controlled study to be conducted within a realistic timeframe, we carefully selected the following two open-source biomedical datasets, containing approximately 15 attributes each, representing typical biomedical data such as demographics, lab measurements, and clinical observations. 

\noindent \textit{\textbf{Task 1: GDC}} In this task, we employ a multi-omics dataset by Li et al.\cite{Li-et-al}, which focuses on the proteogenomic characterization of rare kidney tumors, as our source data. After cleaning and curating the dataset, 18 attributes were retained for mapping. The target dataset is derived from the CPTAC-3 cohort available on the GDC portal, comprising 479 attributes\cite{gdc-site}. 
% \cite{Li-et-al}

\noindent \textit{\textbf{Task 2: PDC}} In this task, we use a proteomic analysis dataset from Woldmar et al.'s study accessed via PubMed ~\cite{WOLDMAR2023100741}. The dataset was refined to 15 matchable attributes. The target schema is PDC v3.0.0, which contains 452 attributes ~\cite{pdc}.

The above two chosen datasets are representative of the kind of data integration tasks commonly performed in biomedical research settings, while being accessible to all participants regardless of their specific subdomain expertise. The participants were required to identify correspondences between attributes in source and target schemas and validate proposed matches. Both tasks were carefully selected to have similar complexity levels for a fair comparison between conditions.

\subsection{Study Procedure}
We conducted the study in a hybrid format allowing for both remote and in-person participation (4 participating in in-person sessions and 8 participating in remote sessions). For remote participants, the study was conducted via Zoom, with access to \toolname provided through a   Kubernetes cluster. In-person participants used university-provided computers on campus.  Regardless of format, all participants had access to their preferred tools (Excel, Python, or R) for the traditional condition to ensure they could work in their familiar environment.

% To create a controlled yet realistic evaluation scenario, we carefully selected open-source biomedical datasets that would be used consistently across all participants. These datasets contained typical biomedical attributes and values, including patient demographics, laboratory measurements, and clinical observations, with approximately 15 attributes per dataset curated by biomedical experts. The datasets were chosen to be representative of the kind of data integration tasks commonly performed in biomedical research settings, while being accessible to all participants regardless of their specific subdomain expertise.
\vspace{2 pt}
\noindent \textbf{Pre-study stage:} 
% Our study procedure was designed to balance thorough evaluation with participant time constraints, creating a streamlined yet comprehensive assessment. 
The participants started with completing a demographic questionnaire about their experience with schema matching and biomedical data. They then received a 5-minute tutorial demo of \toolname's interface use and key features. 
This orientation enabled participants to engage with the tool while still preserving the authentic first-time user experience essential for usability assessment.
% This information helped contextualize their feedback and performance in relation to their background.

% Before using \toolname, participants received a 5-minute introductory tutorial demonstrating the system's interface use and key features. This orientation aimed to enable participants to engage with the tool while still preserving the authentic first-time user experience that is valuable for usability assessment.

\vspace{3 pt}

\noindent \textbf{Study stage:} During the study, the participants were randomly assigned one of the two conditions (baseline/control condition or the test condition using \toolname) first to perform schema matching and after a break, assigned the other condition. Each schema matching task was allocated approximately 20 minutes, providing sufficient time for meaningful engagement while respecting participants' time commitments. We incorporated a 5-minute break between tasks to reduce fatigue and minimize potential carryover effects. The order of study tasks was counterbalanced to compensate for the order-effect. 
In both conditions, participants were asked to: (i) Review the source and target schemas to understand their structure; (ii) Identify potential attribute matches based on names, descriptions, and data types; (iii) Validate or adjust suggested matches using available information; and (iv) Finalize the mapping and generate a list of source attributes and corresponding target attributes.

\vspace{3 pt}

\noindent \textbf{Post-study data collection and feedback stage:} At the end of the study, we conducted semi-structured interviews focused on participant experiences, challenges encountered, and suggestions for improving the tool. To capture the full context of user interactions, we also recorded screen activities (with participant consent) during the sessions. These recordings proved valuable to observe and analyze interaction patterns and the transcripts helped identify specific usability issues that might not be apparent from metrics alone.

\subsection{Analysis}
To evaluate how \toolname{} affects the schema matching processes in biomedical workflows, we employed a mixed-methods approach combining the following:

% \vspace{3 pt}

\noindent \textbf{Assessment Metrics:} We used \textit{NASA Task Load Index (NASA-TLX)} completed by the participants after each task. This test measured the cognitive workload across six dimensions and helped assess how \toolname{} impacted the mental demands of schema matching tasks. Next, to measure the {System Usability Score (SUS)} that measured \toolname's usability, participants filled a questionnaire after completing both tasks. We also measured the participants' performance on metics like schema matching accuracy, number of finished/correct matches and task completion time. The correct matches are evaluated against a ground truth established by domain experts.

\noindent \textbf{Analysis Methods:} 
We conducted statistical analyses using non-parametric methods (Mann-Whitney U test) due to our sample size constraints ($N=12$) and because normality assumptions could not be verified for our data. 
% Specifically, we employed the Mann-Whitney U test on all the collected metrics.
% For comparing performance metrics (accuracy, finished and correct matches, completion time) and NASA TLX scores between conditions, we employed the Mann-Whitney U test, which is appropriate for ordinal data and does not require assumptions about the underlying distribution. 
Effect sizes for these comparisons were calculated using the rank-biserial correlation coefficient ($r$), with interpretations following Cohen's guidelines: small ($0.1 \leq r < 0.3$), medium ($0.3 \leq r < 0.5$), and large ($r \geq 0.5$).
For task completion times, we supplemented the Mann-Whitney U test with survival analysis because completion times were right-censored at the 20-minute maximum time limit. The test employed Kaplan-Meier estimation and log-rank statistic.
% with Kaplan-Meier estimation and log-rank tests to account for right-censored data from our 20-minute task limit. This approach was selected because traditional methods cannot properly handle time-capped observations, potentially biasing results when comparing completion efficiency between conditions.
% For SUS data, which uses Likert-scale responses, we again employed the Mann-Whitney U test to identify significant differences in user perceptions between \toolname~and the baseline condition.
All statistical tests were conducted with $\alpha = 0.05$ as the threshold for statistical significance. We report means ($M$), standard deviations ($SD$), test statistics ($U$, $z$), $p$-values, and effect sizes ($r$) for each comparison.

\begin{figure}[h!]
    \centering
    \includegraphics[alt={NASA Task Load Index (TLX) Comparison Between BDIViz and Baseline. Box plots show workload scores (0-100) across 5 dimensions.}, width= .9\linewidth]{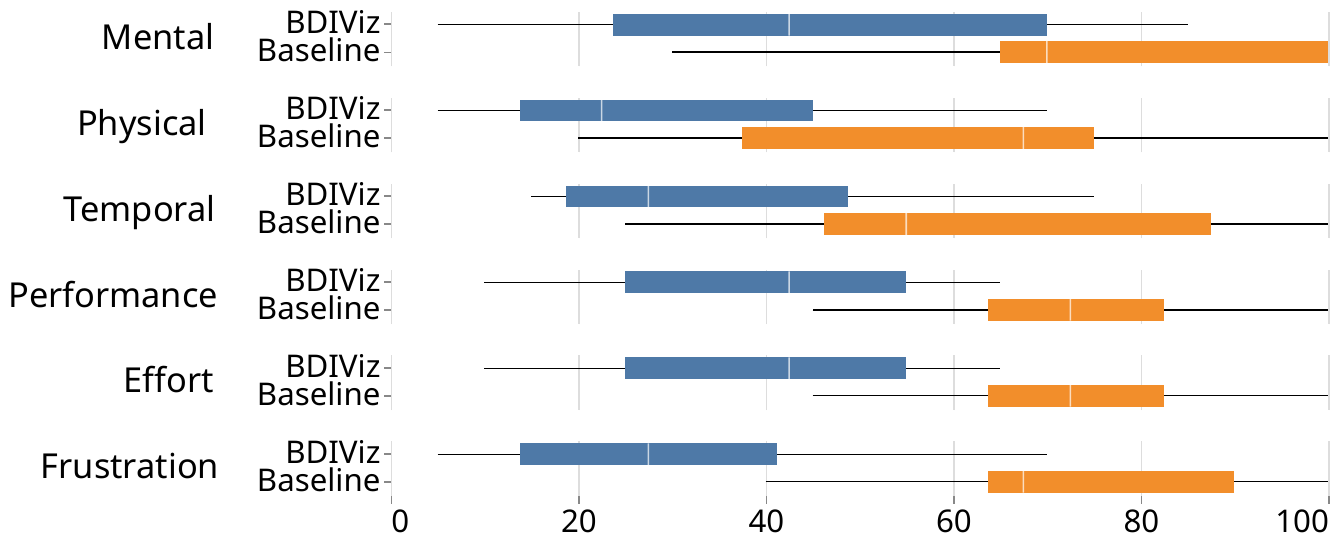}
        \vspace{-.3cm}
    \caption{NASA Task Load Index (TLX) Comparison Between \toolname~and Baseline. Box plots show workload scores (0-100) across 5 dimensions.}
    \label{fig:nasa-tlx}
    \vspace{-.6cm}
    
\end{figure}

\subsection{Quantitative Results}

\begin{figure}[h!]
    \centering
    \includegraphics[alt={Scatter plot comparing the average matching time and accuracy of BDIViz versus Manual methods across two tasks. The x-axis shows the average time (in minutes) per finished matching, while the circle size represents the correct rate (accuracy). Larger markers indicate higher accuracy, demonstrating that BDIViz consistently outperforms Manual curation by achieving faster and more accurate schema matching.}, width = 0.5\textwidth, trim=5pt 5pt 2pt 0pt, clip]{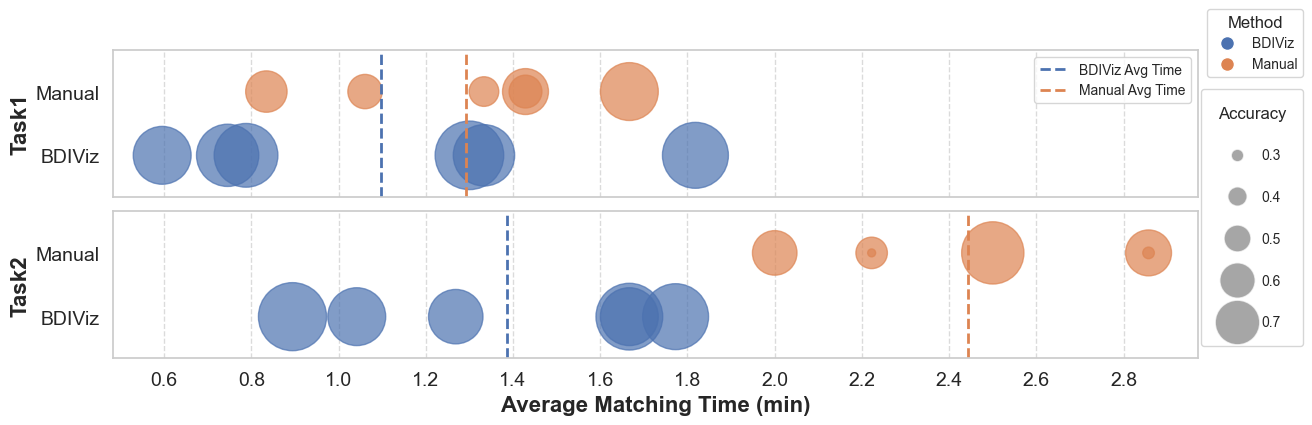}
        \vspace{-.6cm}
    \caption{Scatter plot comparing the average matching time and accuracy of \toolname versus Manual methods across two tasks. The x-axis shows the average time (in minutes) per finished matching, while the circle size represents the correct rate (accuracy). Larger markers indicate higher accuracy, demonstrating that BDIViz consistently outperforms Manual curation by achieving faster and more accurate schema matching.}
    \vspace{-.3cm}
    \label{fig:quantative}
    \vspace{-.15cm}
\end{figure}

 The NASA TLX analysis (Fig.~\ref{fig:nasa-tlx}) demonstrates that \toolname~significantly reduces the overall cognitive workload compared to the baseline approach ($M=43.50$, $SD=18.28$ vs. $M=71.42$, $SD=16.12$, $U=18$, $z=-3.12$, $p=0.002$, $r=0.64$). \toolname~particularly excels at reducing frustration, physical demand, mental demand, and the required effort to complete tasks. Participants also reported reduced temporal demand when using \toolname. 
 Performance measures reveal that \toolname~significantly improves schema matching effectiveness across multiple metrics. The accuracy of matches was substantially higher with \toolname~($M=87.73\%$, $SD=4.41\%$) compared to the baseline approach ($M=58.19\%$, $SD=18.17\%$), representing a $50.77\%$ improvement.
 In addition, participants derived more correct matches while completing the task much faster 
 %also achieved significantly more correct matches using \toolname while needing significantly less time for task completion 
 (Log-rank test: $\chi^2=6.12$, $p=0.013$) (Fig.~\ref{fig:quantative}). 
 \revision{
 We saw greater improvements in performance on Task 2, which presents a higher level of ambiguity, making it more challenging than Task 1.
 %naming styles. 
 This aligns with our goal of supporting users in complex tasks, that requires them to explore matches in greater detail, compare attribute values, and obtain additional context to make a decision.
 %less interpretable scenarios. 
 We expect that \toolname would yield even bigger gains for larger and more challenging schema matching tasks than the ones in our user study. 
 % However, due to time constraints we limited the evaluation to smaller, representative datasets. 
 Our final system usability scale (SUS) score for \toolname was 71.87} showing that \toolname~achieved significantly better user ratings among all questions. \revision{ Detailed SUS results and analysis are provided in the supplementary material.}

\textbf{In summary,} these results demonstrate that \toolname~enables users to perform schema matching tasks with substantially higher accuracy and efficiency compared to conventional approaches.

\subsection{Qualitative Results} % \dt{maybe point to Rs/DCs}}
We conducted a thematic analysis of interview transcripts and observational notes, with two researchers independently coding and refining themes through consensus. This analysis identified user experience patterns, challenges, and comparative insights on \toolname versus traditional methods, linking emerging themes to quantitative findings for a comprehensive understanding of \toolname’s impact on schema matching.

\para{Identification of easy matches saved time to focus on critical matches} User feedback consistently highlighted the value of \toolname’s strategy to prioritize straightforward matches, enabling experts to quickly validate high-confidence candidates while focusing on ambiguous cases. The interactive heatmap, with its clear color intensity indicating matching confidence, allows users to easily identify matches that need closer examination. As participant 10 (p10) noted, \participantquote{(\toolname) really helps me avoid hesitation about where to start. I can just begin with what I've suggested.} Similarly, p2 commented that they \participantquote{like that the heatmap visualization—it [gave] an overview of the matching score while providing detailed comparisons, and the interface is smooth and intuitive.} This demonstrates how \toolname contrasts with traditional methods, where all potential matches demand equal scrutiny, and helps users allocate their cognitive resources more efficiently.

% \subsubsection{Interactive exploration helped nuanced decision making for critical matches}

% The ability to interactively explore data distributions and metadata through \toolname significantly enhanced participants' decision-making process for ambiguous attribute matches. Participants particularly valued the expandable nodes feature, which revealed detailed value distributions upon interaction. For example, Ey pointed out, \participantquote{I especially like the heat visualization of the matching, which gives you an overview of the matching score while provides detailed comparison.}.
% When working with complex biomedical attributes that had similar names or structures, participants utilized these interactive elements to examine the underlying data characteristics before making matching decisions. \participantquote{[TBD]}. 
% Several participants demonstrated this benefit during the study sessions by initially expressing uncertainty about a match, then gaining confidence after exploring distribution visualizations that revealed pattern similarities between attributes. For example, Ex mentioned that \participantquote{My workflow typically involves selecting the highest value, then clicking to check the matching patterns of attributes. This allows me to see the distribution of each attribute, which is incredibly helpful.}.

\para{Interactive exploration helped nuanced decision making for critical matches} Participants valued interactive features such as expandable nodes that reveal detailed value distributions. P2 noted that they \participantquote{especially like the heat visualization—it provides an overview of the matching score along with detailed comparisons.} When working with complex biomedical attributes that have similar names, these interactive elements allowed users to examine underlying data characteristics for decision making. P10 explained, because their \participantquote{workflow involved selecting the highest value, then clicking to check the matching patterns, seeing each attribute's distributions was incredibly helpful.}

\para{LLMs helped with validation} In \toolname, the LLM generates visually encoded explanations for each matching candidate, allowing users to assess and validate recommendations. The explanations use intuitive icons and color cues to indicate the type of reasoning and suggested decision, aligning with users' thought processes. Participants appreciated this clarity; p10 remarked, \participantquote{The second best feature is the agent... They provide the rationale behind each match, offering variant reasoning tracks that align with my own thought process.} These explanations proved especially useful when users encountered nuanced terminology or attribute names, helping them verify challenging matches. As p2  noted, \participantquote{I refer to the agent's suggestions and explanations often—I feel more informed, confident, and comfortable using \toolname for schema matching.} Overall, the LLM-driven validation not only highlighted considerations that might otherwise be overlooked but also increased user confidence in the final matching decisions.

% \subsubsection{Extra information helped in knowledge discovery}

% Biomedical datasets often contain specialized terminology that not all users are familiar with. Instead of having to interrupt their workflow to search external resources—as they reported doing with traditional tools—participants could access relevant domain knowledge directly within the interface. For example, Ex mentioned that \participantquote{without \toolname's help, I have to google and use different tools, which is time-consuiming}. This integration of reference information within the matching workflow appeared to both improve matching accuracy and reduce the time spent on the task, which is also reflected in our quantitative analysis.

\para{Extra information helped in knowledge discovery} Biomedical datasets often contain specialized terminology that may be unfamiliar to users. Rather than interrupting their workflow to consult external resources—as is common with expert workflows—participants appreciated having relevant domain knowledge integrated directly into the interface. P5 noted, \participantquote{Without \toolname's help, I have to google and use different tools, which is time-consuming.} This integration not only improved matching accuracy but also reduced the overall time spent on the task, as reflected in our quantitative analysis.

\textbf{In summary}, the results demonstrate \toolname's capabilities to address the design requirements as well the design considerations (\ref{sec:co-design}) for enabling effective and efficient domain-specific schema matching.  
% \newpage
\section{Expert Case Study Applications} 
\label{sec:case-study}

% \claudio{Would it be possible to have figures that highlight these case studies? The paper is missing figures in the last several pages.}

% \jf{Should this be moved to the end?}
To further demonstrate how \toolname generalizes for multiple real-world use cases while also accommodating for large and complex biomedical schemas, we present two use cases provided by our expert co-designers. 
% \jf{We should add the use cases as a 3rd subsection of the evaluation}

% \jf{The titles for the case study should describe the actual task; in the text, we can highlight the useful features of bdiviz and how they simplified/streamlined the task}

%\subsection{Case Study 1: Using \toolname for Large-scale Dataset-to-Schema Matching Task for CPTAC}
\subsection{Using \toolname for Large-Scale Dataset-to-Schema Matching Tasks for CPTAC}

\begin{figure}[t]
    \centering
    \includegraphics[alt={Case Study 1: Heatmap visualization of curated matching candidates for Dou et al.'s endometrial carcinoma dataset to the GDC schema in the CPTAC-3 study.}, width= \linewidth, trim=100pt 120pt 100pt 120pt, clip]{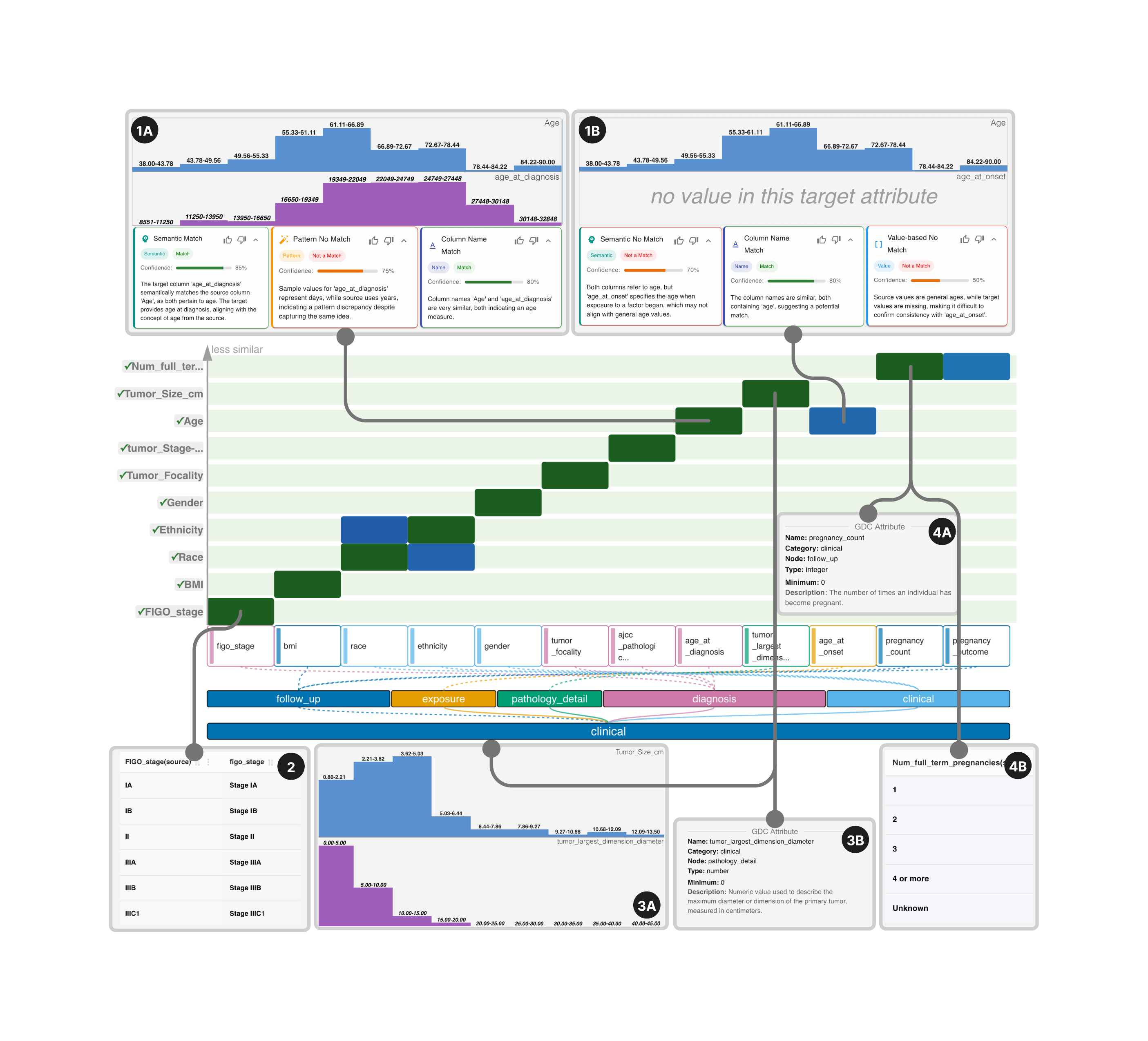}
    \caption{Case Study 1: Heatmap visualization of curated matching candidates for Dou et al.'s endometrial carcinoma dataset to the GDC schema in the CPTAC-3 study.}
    \label{fig:case-study}
    \vspace{-.5cm}
\end{figure}

We used \toolname to harmonize a dataset generated in the context of the Clinical Proteomic Tumor Analysis Consortium (CPTAC) with the GDC schema.
%a large-scale source dataset to a target schema from the Clinical Proteomic Tumor Analysis Consortium (CPTAC). The CPTAC initiative seeks to make data from multiple studies available in an integrated, standardized format within the Cancer Research Data Commons (CRDC)\cite{CRDC,Hinkson2017}.
The source dataset (referred as Dou) includes data for patients from an endometrial carcinoma study and consists of 179 attributes~\cite{Dou-source}. 
%was mapped to the GDC schema with over 700 attributes. 
Through a manual process, biomedical experts identified correspondences between 19 attributes in Dou and GDC elements~\cite{liu2024magnetocombiningsmalllarge}.
%The dataset, previously curated by biomedical experts, was mapped to 19 corresponding attributes (\textit{ground truth}) in the GDC schema using a combination of manual methods and automatic matchers.
% This case study demonstrates the workflow of using \toolname to curate a large-scale source dataset into a target schema derived from the Clinical Proteomic Tumor Analysis Consortium (CPTAC) program. The CPTAC initiative seeks to make data from multiple studies available in an integrated, standardized format within the Cancer Research Data Commons (CRDC)\cite{CRDC,Hinkson2017}. Patient case and sample data from these studies are often published in supplemental tables of scientific papers, but they typically exhibit heterogeneous variable names and value types. The source dataset used in this case study is drawn from Dou et al.'s work on endometrial carcinoma, comprising 179 attributes~\cite{Dou-source}. The target schema is based on the GDC, which contains over 700 attributes; we retrieved the complete study cohort from CPTAC-3 to capture real data distributions. This dataset was previously curated by a team of biomedical experts using a combination of manual methods and automatic matchers, resulting in 19 source attributes that were successfully mapped to corresponding attributes in GDC~\cite{liu2024magneto}.

%\jf{This reads a lot like a manual -- the main takeaway is that it is easy to verify and surprisingly, additional mappings were found}
Using \toolname, within 15 minutes,  a biomedical researcher was able to verify the 19 matches as well as identify 10 additional matches that had been missed in the manual schema mapping process. The user reported that it was easy to validate the accepted matches, which included \emph{easy} matches, for example the attributes \texttt{FIGO\_stage} and \texttt{figo\_stage}, whose values could be easily aligned (e.g., “IA” to “Stage IA”, “IB” to “Stage IB”) (\figref{\ref{fig:case-study}}{2}).
To assess matches with numerical attributes, they leveraged the GDC attribute description and the comparisons of value distributions (\figref{\ref{fig:case-study}}{3A, 3B}).
%by leveraging attribute name and value similarities. It takes our server \textbf{1.5 minutes} to generate matching candidates and the user able to complete the matching task in \textbf{15 minutes}. \textbf{6 out of 19 \textit{ground truth}} were easy matches. For example, \texttt{FIGO\_stage} and \texttt{figo\_stage} was accepted as an exact match, with the attribute names differing only in case and the corresponding values aligning (e.g., “IA” to “Stage IA”, “IB” to “Stage IB”). For attributes like \texttt{Tumor\_Size\_cm}, we validated matches based on value distributions and GDC descriptions. Despite mixed feedback, the system accepted \texttt{age\_at\_diagnosis} as a valid match, demonstrating the system’s ability to utilize domain knowledge. 

%Our system successfully identified additional ten \textit{ground truth} mappings other than the 19 pairs, 
Among the newly-identified mappings was the match from \texttt{Num\_full\_term\_pregnancies} to \texttt{pregnancy\_count}. The expert verified that the unique values for the source attribute (\texttt{1}, \texttt{2}, \texttt{3}, \texttt{4 or more}) was consistent with the GDC description, "The number of times an individual has become pregnant"(\figref{\ref{fig:case-study}}{4A, 4B}). This highlights the ability of \toolname to \emph{address scalability challenges,} by matching schemas effectively even for attributes distributed across large datasets.
When faced with similar candidate pairs, such as \texttt{age\_at\_diagnosis} versus \texttt{age\_at\_onset}, the expert relied on detailed agent explanations to differentiate between them. The system rejected \texttt{age\_at\_onset} due to its specific reference to the onset of exposure, highlighting \toolname's ability in \emph{facilitating nuanced decision-making} (\figref{\ref{fig:case-study}}{1A, 1B}).
%
% The pattern matching supported by the agent also facilitated the mapping of \texttt{WXS\_normal\_sample\_type} to \texttt{sample\_type} by identifying that the source value \texttt{Blood\_normal} corresponded with target values such as \texttt{Blood Derived Normal}. 

% \jf{I don't understand this}
% Furthermore, leveraging the vertical attribute comparison feature—by adjusting the Similar Attribute slider—we aggregated related attributes (\texttt{WGS\_tumor\_sample\_type} and \texttt{WXS\_tumor\_sample\_type}) and successfully mapped them to \texttt{sample\_type}. 

\subsection{Using \toolname for Dataset-to-Dataset Matching for LUAD Studies}
%Address Nuanced Variations in Joining LUAD Studies}
% \subsection{Integrating Large Scale Datasets}

An expert used \toolname to integrate datasets from two distinct lung adenocarcinoma (LUAD) studies published on PubMed. The source dataset from Xu et al. comprises 40 attributes~\cite{XU2020245}, while the target dataset from Gillette et al. contains 74 attributes~\cite{GILLETTE2020200}. Within 10 minutes, the expert identified 13 matches; this is in contrast to the 8 attributes previously identified through manual matching.
%—5 more than those found in the previously curated harmonized table using \toolname \jf{were the original 8 matches also derived by bdiviz?}.

% \footnote{https://www.sciencedirect.com/science/article/pii/S0092867420306760}
% \footnote{https://pubmed.ncbi.nlm.nih.gov/32649874}

% ~\cite{XU2020245}
% ~\cite{GILLETTE2020200}
%
\toolname\ effectively addresses the nuanced semantics common in biomedical attributes.  For example, the source attributes \texttt{Tumor Stage (A/B)} and \texttt{Tumor stage} received high similarity scores with the target attribute \texttt{Stage}. By leveraging vertical comparison, i.e., comparing the similarities between the target attribute and the source attributes, and inspecting their value distributions, the expert observed that \texttt{Tumor Stage (A/B)} exhibited peaks at \texttt{IB} and \texttt{IIIA} (corresponding to \texttt{1B} and \texttt{3A} from \texttt{Stage}), whereas \texttt{Tumor stage} showed peaks at \texttt{T2aN0M0} and \texttt{T2aN2M0}, indicating a mismatch. This example underscores \toolname's capability to \emph{handle subtle attribute differences}. % during dataset integration.

%\textbf{In Summary,} 
The real-world case studies show that \toolname is effective at streamlining different schema matching tasks for diverse biomedical datasets and schemas. They also confirm the findings in our user study in that not only does \toolname makes users more efficient, but it also leads to mappings that are more accurate and complete.
\section{Discussion}
%- \dt{maybe point back to the DCs here too?}}
While we focus on biomedicine, our approach has broader relevance to the HCI and Visualization communities, as discussed below.

% \vspace{5 pt}

% \subsection{Contributions:}
\para{Designing schema matching tools for data-intensive domains} \toolname tackles the challenges of schema matching in data-intensive domains by integrating interactive heatmaps, value comparisons, and analysis panels that bridge automated matching with expert curation. \clarification{This design not only benefits biomedical data—with its complex terminologies and large attribute sets—but also applies to other fields such as urban science, that requires the integration of diverse and complex datasets.}
%such as legal compliance, where harmonizing diverse documents is critical.
% JF: our approach applies to structure data not to unstructured documents
% This work addresses the challenges of schema matching in data-intensive domains like biomedicine. \toolname tackles scalability issues in large, complex biomedical datasets where traditional methods fall short. By combining interactive heatmaps, value comparisons, and detailed analysis panels, \toolname bridges the gap between automated schema matching and expert-driven curation. These techniques extend beyond biomedicine to domains like legal and regulatory compliance, where firms must harmonize legal documents, contracts, and compliance records across jurisdictions. With evolving regulations and complex terminologies, robust tools similar to \toolname are ideal for legal experts, helping them efficiently match and integrate regulatory schemas with greater accuracy. 

% \vspace{5 pt}

\para{Designing visualization tools to empower domain experts} \toolname empowers domain experts to make informed, data-driven decisions 
%without deep machine learning expertise 
by incorporating a user-in-the-loop methodology. 
%This iterative refinement allows experts to focus on nuances while quickly validating high-confidence matches. 
Informed by formative studies and co-design sessions, our design aligns closely with natural expert workflows, ensuring that detailed visualizations—such as the heatmap, attribute distributions and value comparisons—support rather than disrupt the matching process.

% \toolname also showcases the importance of designing tools that empower domain experts to make data-driven decisions without requiring a deep expertise in machine learning methods. We specifically designed \toolname with the the expert user-in-the-loop methodology, allowing users to iteratively refine matches suggested by our system. This approach ensures that experts always remain in control of the decision-making process, as they can accept, reject, or refine candidate matches based on their domain knowledge and the visualized data distributions. For biomedical data harmonization, this proved essential as automated methods can fall short due to the domain-specific data complexity.

% \vspace{5 pt}

\para{Studying expert workflows to design tools} 
Our formative study and co-design sessions with experts underscore the importance of understanding user workflows prior to tool development. By aligning \toolname\ with the real-world practices of domain experts, we have created a system that complements natural workflows. This user-centered design ensured that the tool is not only usable but also effective, a principle that should guide the development of future domain-specific tools.

% Another key takeaway of our work is the importance of understanding expert workflows before developing technical tools. Our formative study and codesign sessions with biomedical experts informed the design of \toolname, ensuring that the tool aligns with the real-world needs of these users. Tools for domain-specific tasks should be grounded in how experts approach problems, how they use the data, and what steps they take to refine and verify their work. By incorporating features like detailed value comparisons and distributions, we built a tool that complements the natural workflows of domain experts, rather than imposing a rigid, predefined structure on them. Future tools for other domains should similarly begin with a deep understanding of user workflows and decision-making processes to ensure that they are not only usable but truly useful.

% \vspace{5 pt}

\para{Using LLMs and Visualization for Schema Matching} 
Our work demonstrates how combining LLMs with interactive visualization enhances schema matching, particularly in complex domains like biomedicine. LLMs offer strong semantic reasoning, but ambiguity, domain-specific terms, and subtle value differences still demand expert oversight. Visualization enables users to explore, validate, and refine matches, ensuring alignment with domain knowledge. Even as LLMs improve, \toolname will remain essential for expert validation, adapting to custom attributes, and supporting user corrections. This human-in-the-loop approach ensures adaptability to new datasets, evolving schemas, and unforeseen edge cases, making schema matching more transparent, interpretable, and trustworthy for diverse applications.

% Our work highlights the synergy between LLMs and interactive visualization in enhancing schema matching tasks. LLMs provide powerful semantic reasoning capabilities, enabling automated match suggestions based on linguistic patterns, contextual understanding, and learned domain knowledge. However, schema matching remains a complex task, especially fields like biomedicine, where attributes often have ambiguous names, use domain-specific terminology, and present subtle value differences. Visualization plays a crucial role in bridging this gap by allowing users to interactively explore, validate, and refine matches, ensuring that automated suggestions align with domain expertise. While LLMs are evolving rapidly and may eventually achieve near-perfect matching accuracy, \toolname remains valuable by offering expert-driven validation, handling custom attribute names, and supporting user corrections. This human-in-the-loop approach ensures adaptability to new datasets, evolving schemas, and unforeseen edge cases, making schema matching more transparent, interpretable, and trustworthy across diverse applications.

\section{Limitations and Future Work}  
Our evaluation and case studies have demonstrated \toolname's effectiveness in streamlining schema matching tasks and improving accuracy. Nonetheless, the system has limitations that present opportunities for future research.
First, the performance of \toolname depends on the effectiveness of the underlying matching methods. Although we mitigate this limitation by 
combining an ensemble approach
%combining multiple matching methods and employing 
with LLM-based validation, it is possible that all methods fail to identify correct matches, requiring users to resort to manual identification. 
Future work will investigate extensions to \toolname that better assist users in this manual discovery process, particularly through search mechanisms that help identify false negative matches that automated methods miss entirely.

% Second, our current implementation provides LLM interaction through fixed UI components with predefined queries. A promising direction for future work is the integration of a conversational interface that would allow users to formulate task-specific questions about potential matches, enabling more flexible and targeted exploration of semantic relationships between attributes. 
%This approach could significantly enhance users' ability to resolve complex matching decisions by providing contextual information beyond what fixed UI components can offer.

\revision{
Second, although our heatmap visualization and interactive filters improves scalability over node-link approaches, it could still become visually and cognitively demanding when dealing with very large schemas. 
% I find it improbable that there are schemas with thousands of attributes
%containing hundreds or thousands of attributes, 
As the system is deployed, we plan to explore additional abstractions and interactive exploration techniques. For example, some users have pointed out that they would like to be able to also use a tree-based representation for the source table.
}

Third, the quality of LLM-derived validations and explanations inherently depends on the model's knowledge and reasoning capabilities. Since these models can make errors, their limitations will inevitably be reflected in the information provided by \toolname. However, with the rapid advances in language model development, we anticipate that these capabilities and reliability will continue to improve, enhancing the system's performance without architectural changes.

Fourth, we designed \toolname to support matching from dataset to dataset and from dataset to a schema. Another important scenario involves the harmonization of multiple datasets to a schema, i.e., data from multiple sources is required to populate the schema. 

Finally, our evaluation demonstrated the effectiveness and strengths of \toolname for schema matching in the biomedical domain. \clarification{While the tool was designed to be general and support arbitrary schemas and datasets, further evaluation in different domains would provide valuable insights into the adaptability of the tool and potential refinements.}

\section{Conclusion}
% \revision{We introduced BDIViz, a visual analytics system for biomedical schema matching that combines ensemble-based algorithms, interactive visualizations, and LLM-based validation. Informed by formative studies with domain experts, BDIViz addresses key usability and accuracy challenges in schema matching. Our evaluation showed that it improves matching accuracy while reducing cognitive load and curation time. Case studies demonstrated its effectiveness across diverse real-world tasks.}
% With biomedical datasets growing in diversity and volume, effective schema matching tools are increasingly critical. 
% BDIViz is a key step in addressing this need by transforming the traditionally manual and error-prone matching process into a guided interactive workflow combining human expertise with computational capabilities. By streamlining curation and improving harmonization accuracy, BDIViz contributes to making data interoperable and enabling analyses that can fuel scientific discoveries and advance biomedical research.

%old conclusion (submission version)
We presented BDIViz, a visual analytics system that addresses critical challenges in biomedical schema matching through user-centered design. Our formative studies with domain experts informed an approach that combines ensemble-based matching algorithms, interactive visualizations, and LLM-based validation. Evaluation demonstrated that BDIViz significantly improves matching accuracy while reducing cognitive load and curation time. Case studies confirmed the system's effectiveness across diverse real-world biomedical schema matching tasks.

As biomedical datasets continue to grow in diversity and volume, effective schema matching tools become increasingly critical for data integration. 
BDIViz transforms the traditionally manual and error-prone matching process into a guided interactive workflow that combines human expertise with computational capabilities. By improving harmonization accuracy and streamlining curation, BDIViz contributes to making biomedical data more interoperable, enabling analyses that advance scientific discovery and biomedical research.

\acknowledgments{
This work was supported by NSF awards IIS-2106888 and OAC-2411221, and the DARPA ASKEM program
Agreement No. HR0011262087. The views, opinions, and findings expressed are those of the authors and should not be interpreted as representing the official views or policies of the DARPA, the U.S. Government, or NSF. 
% We thank the biomedical experts for their participation, contributions, and feedback in the co-design sessions. 
}
\balance
% \newpage

\bibliographystyle{abbrv-doi-hyperref}

\bibliography{bdiviz,bdiviz-revision}

\appendix

\section{Supplementary Material}
Please refer to the supplementary material for (i) detailed background information on the experts interviewed during the formative study, (ii) the full LLM prompt used in the system, and (iii) complete scores and analyses of the System Usability Scale (SUS) results across multiple dimensions. 

%%MOVE THE FOLLOWING TO THE SUPPLEMENTARY MATERIALS
% \section{Supplementary Material}
% \subsection{Expert Participants}
% \label{sec:experts}

% \revision{To design a tool that aligns with biomedical researchers' data harmonization workflows,
% we conducted a formative study with biomedical experts and data scientists. We recruited five expert participants (Table \ref{tab:experts}) from our collaborator networks. Among these experts, three are biomedical researchers and data scientists from a major academic medical institution (E1 - E3), one is a principal scientist from a major research organization (E4), and one is a data librarian at a prominent academic medical library (E5).} 

% % \vspace{4 pt}
% \input{tables/experts}
% % \vspace{-10 pt}

% \subsection{System Usability Score Details}

%  \revision{The final system usability scale (SUS) score for \toolname was 71.87 and further details on the SUS survey
%  % Further details of the quantitative analysis results 
% are shown in Table~\ref{tab:sus} illustrating that \toolname~achieved significantly better user ratings among all questions. }

%  \input{tables/sus}

\end{document}